\documentclass[a4paper,11pt]{article} \pdfoutput=1 
\usepackage{jheppub} % for details on the use of the package, please % see the
\usepackage{chay,graphicx,amsmath}

\newcommand{\euv}{\epsilon_{\mathrm{UV}}}
\newcommand{\eir}{\epsilon_{\mathrm{IR}}}

\newcommand{\dfl}{\frac{d^D l}{(2\pi)^D}}
\newcommand{\dimf}{\Bigl(\frac{\mu^2 e^{\gamma_{\mathrm{E}}}}{4\pi}\Bigr)^{\epsilon}}
\newcommand{\lone}{\Bigl( \frac{\ln (1-z)}{1-z}\Bigr)_+}
\newcommand{\one}{$\mathrm{SCET}_{\mathrm{I}}$}
\newcommand{\two}{$\mathrm{SCET}_{\mathrm{II}}$}
\newcommand{\cy}{\mathcal{Y}}
\title{\boldmath Proper factorization theorems in high-energy scattering  near the endpoint}

\author[a]{Junegone Chay} 
\author[b]{Chul Kim} 
\affiliation[a]{Department of
Physics, Korea University, Seoul 136-713, Korea} 
\affiliation[b]{Institute of
Convergence Fundamental Studies \& School of Liberal Arts, Seoul National
University of Science and Technology, Seoul 139-743, Korea}
\emailAdd{chay@korea.ac.kr} 
\emailAdd{chul@seoultech.ac.kr}

\abstract{ Consistent factorization theorems in high-energy scattering near
the threshold are presented in the framework of
the soft-collinear effective theory. Traditional factorization theorem
separates the soft and collinear parts successfully, but   a final step should be supplemented
if each part encounters infrared divergence.  
We present factorization theorems in which the infrared divergences appear only in the
parton distribution functions and the infrared divergence is removed by carefully separating and
reorganizing collinear and soft parts. The underlying physical idea is to 
isolate and remove the soft contributions systematically from the collinear part
in loop corrections order by order. After this procedure, each factorized term 
in the scattering  cross sections is free of
infrared divergence, and can be safely computed using perturbation theory. This
factorization procedure can be applied to various high-energy scattering
processes. We show factorization theorems in Drell-Yan processes, deep
inelastic scattering and Higgs production near the  endpoint.}

\begin{document} 
\maketitle 
\flushbottom %\baselineskip 3.0 ex
\section{Introduction} Theoretical predictions on high-energy scattering are
based on the factorization theorem, in which the cross section is factorized
into the hard, collinear and soft parts. It means that, though the strong
interaction affects the scattering processes at different scales in various stages, 
these effects can
be separated according to their kinematic regimes. The
hard part consists of the partonic cross sections with hard momenta. The
collinear part describes the effects of the collimated energetic particles. For hadron
colliders, the collinear part involves the parton distribution functions (PDF)
in the initial state. If an exclusive process is considered, say, a pion production, there
appears the fragmentation function which depicts the hadronization process
from a parton into a hadron. This is also classified as the collinear part. The soft part
describes soft gluon exchanges among different collinear sectors. These three
parts are decoupled and there is no communication among them. Therefore it is
possible to compute the contributions for each part separately
using perturbative quantum chromodynamics (QCD).

The factorization property in various high-energy scattering processes has
been well established in full QCD \cite{Sterman:1986aj,Catani:1989ne}.
In full QCD, once the scattering cross sections are written, they are examined 
closely by dividing collinear, soft momentum regions. A laborious  analysis 
enables the  factorization proof that collinear and soft parts are decoupled.
The factorization proof becomes more elaborate in
soft-collinear effective theory (SCET)~\cite{Bauer:2000ew,Bauer:2000yr,Bauer:2001yt} since  the separation of collinear and soft parts is established from the outset at the operator level. 
The scattering cross sections 
are written as a product of the hard coefficient in terms of the Wilson coefficients, and the 
collinear and soft parts, which are expressed as the matrix elements of gauge-invariant 
operators.   The factorization of the hard, collinear and soft
parts in the hard scattering is now on a firm basis  though the collinear and soft
parts may be still infrared divergent in some cases.  The detailed form of the 
factorization formulae may be different in full QCD or in SCET,  but basically 
the scattering cross sections can be written in a factorized form with the hard, collinear and
soft parts. We call this development  the traditional factorization.

Since the advent of Large Hadron Collider (LHC), precise theoretical predictions
on high-energy scattering can be under experimental scrutiny. As well as the
fixed higher-order corrections in the strong coupling $\alpha_s$, the resummed
 inclusive cross sections using the renormalization group technique are available when there are
large threshold logarithms~\cite{Sterman:1986aj,Catani:1989ne, 
Catani:1996yz,Becher:2006nr,Idilbi:2006dg,Becher:2006mr}. 
The theoretical accuracy available for the comparison with experiment can be attained 
by computing the factorized parts at higher orders or resumming large logarithms.
The traditional factorization theorems have been successful because the collinear and 
soft parts can be expressed in terms of  gauge-invariant operators and there is no communication 
between them. However, the  traditional factorization theorems are
plagued by infrared divergences. 

In massless gauge theories like QCD, there exists  
infrared (IR) divergence as well as ultraviolet (UV) divergence.  When a
massless quark emits a gluon, collinear divergence appears when these two
particles are collinear.  Soft divergence  emerges when the energy of
the emitted gluon becomes soft. If the emitted gluon is both soft and collinear
to the quark, soft and collinear divergences occur simultaneously.
This type of divergence is referred to as the IR divergence. 
The UV divergence also shows up, but they can be removed by counter terms. 
The remnant after removing the UV divergence governs the scaling behavior.
On the other hand, the IR divergences cannot be removed by hand. 
Instead, physical quantities should be free of IR divergence.  
The Kinoshita-Lee-Nauenberg (KLN) theorem~\cite{KLN} succinctly
states that the IR divergences appearing in real gluon emissions and in virtual
corrections at any given order in $\alpha_s$ should cancel in scattering cross
sections.

There is no IR divergence in the hard part, and most of the IR divergences are cancelled 
in the sum of the soft and collinear parts. The exception is the radiative correction to the PDF.
But since the PDF is basically a 
nonperturbative quantity, the IR divergence can be absorbed in the PDF. 
However,  as we will explain, the soft and collinear parts in the  traditional factorization theorems
may contain IR divergences, and furthermore mixing between IR and UV divergences.
In this case, the collinear and soft parts themselves cannot be regarded as physical quantities.
The evolution of the collinear and soft parts
does not make sense either because these parts contain
IR divergences and mixed divergences, hence the renormalization scale 
dependence does not stem from UV divergences alone.

Note that the IR divergences are not artifacts in calculating radiative corrections, 
and they always appear irrespective of the regularization
methods. If we use the dimensional regularization to regulate both UV and
IR divergences, the UV (IR) divergences appear as poles of $\euv$ ($\eir$). If
the UV divergence is regulated by the dimensional regularization, and the IR
divergence is regulated by the offshellness of external particles, the UV
divergence still appears as poles in $\euv$, but the IR divergence appears as
the logarithms of the offshellness. Unless the IR divergences are taken care of,
the physical interpretation of each part with IR divergences is not possible.
We need more than the traditional factorization method by employing an additional
step to render each part IR-finite.

Here we confine ourselves to the inclusive scattering cross sections. But it will be
important to see if we can extend the 
methodology employed here to more exclusive physical quantities such as event shape, jet mass,
jet vetos, etc.. 
Many of these exclusive quantities  may be presumably IR finite because there can exist physical
parameters which act as IR cutoff. In inclusive processes, the IR cutoff is missing and if any 
factorized parts contain  IR divergences, it is critical to find the method in which the factorized
parts are IR finite. In fact, this happens in the inclusive scattering: DIS, Drell-Yan process near
the hadronic threshold, and in the Higgs production near the partonic threshold.  Careful 
redefinitions of the soft function are needed to define IR-finite quantities.

Our  idea of factorization respects the philosophy of the  traditional factorization
theorems in the sense that the inclusive cross sections are factorized and the collinear
and soft parts can be expressed in terms of the gauge-invariant operators. But
it goes  beyond earlier threshold factorization theorems by givng proper threshold 
factorization theorems with IR-finite objects.  As a result, 
the IR divergence resides only in the PDF, as in full QCD, and the remaining
parts in the factorization formula are free of IR divergences. Only after this
reorganization, the IR-finite quantities can be computed systematically in
perturbation theory. However, this new factorization is not just
reorganizing the factorized parts to shuffle IR divergences as a matter of
convenience, but there is a deeper
physical reason. In loop calculations, the collinear part unavoidably covers
the kinematic region with soft momentum when the loop momentum becomes soft. 
But the soft region is already accounted for in the soft part. Therefore the soft
contribution in the collinear part should be removed in an appropriate way
to avoid double counting.

This removal can be systematically performed in SCET.  The systematic 
technique to remove the soft contribution consistently 
from the collinear part is called the zero-bin subtraction \cite{Manohar:2006nz} 
in SCET.  The basic idea of the zero-bin subtraction is that, for each collinear Feynman 
diagram, the soft contribution can be computed by taking the limit where the loop 
momentum becomes soft, and then this soft contribution is subtracted from the naive 
collinear loop calculation.
The zero-bin subtraction is not an esoteric technique used in SCET, but a simple tool to 
avoid double counting.  The same idea can also be realized in full QCD. In full QCD, a careful
separation of kinematic regions with different momenta should be performed,  but the 
extraction of the soft contribution from the collinear part is tedious. SCET is an effective 
theory best fit to this purpose. SCET involves all the relevant fields necessary in the factorization, 
hence the factorization proof in SCET is equivalent to that in full QCD  itself, but with more 
transparency.

We emphasize that care should be taken to perform an appropriate zero-bin 
subtraction. Naively any soft limit of the collinear loop momentum may do for calculating 
the zero-bin contributions, but in fact the soft limit of the collinear momentum should be 
determined by the size of the momentum entering the soft part.
This is the essential point in introducing the zero-bin subtraction. If $Q$ is
the large scale in the system, and $\lambda$ is a small parameter in SCET, 
the soft part describes the fluctuations of order $Q\lambda$ or $Q\lambda^2$, 
and the soft limit in the collinear part should also be of the same order. If the soft
part describes the interaction with momentum of order $Q\lambda$ 
($Q\lambda^2$), the soft limit  in the collinear part should also be of order 
$Q\lambda$ ($Q\lambda^2$). If the soft momentum in the 
soft part is of order $Q\lambda$, while the soft limit in 
the collinear part is of order $Q\lambda^2$,  or vice versa, there is a mismatch between 
the soft part and the soft contribution in the collinear part. With this mismatch, 
the zero-bin subtraction does not produce any sensible results, 
and the double counting is not removed completely.

After the double counting problem is handled properly, 
the scattering cross sections near the endpoint again factorize, but 
in a different form. They are factored
into the hard part, the collinear part, and the soft kernel which consists of the 
original soft contributions and the removed soft part from the collinear part 
through the zero-bin subtraction. Then each term except the PDF is IR finite, 
and the resummation of large logarithms can be performed using the
renormalization group equation. We show the factorization theorems in this
picture in deep inelastic scattering (DIS), Drell-Yan (DY) processes near the hadronic endpoint,  
and present the one-loop results explicitly. This factorization proof  
also applies in computing resummation near the partonic threshold in the Higgs 
production, and it forms a robust basis in performing 
a systematic analysis of the threshold resummation.

The structure of the paper is as follows: In Sec.~\ref{factscet}, the features
of SCET are described briefly and the factorization theorems in DY, DIS  
processes near the endpoint are outlined. In Sec.~\ref{dyfac}, 
the detailed factorization theorem is presented for DY processes, and
 the factorization theorem for DIS 
near the hadronic endpoint is shown in Sec.~\ref{disfac}. The factorization theorem for the 
Higgs production near the partonic threshold is presented in Sec.~\ref{higgsfac}. In
Sec.~\ref{softf}, the soft functions for all these processes are computed at one loop 
in terms of the matrix elements of the soft Wilson lines. The soft functions at higher
orders clearly show that they include
IR divergence and mixing between UV and IR divergences, hence not physical in
themselves. In Sec.~\ref{colli}, we introduce
the collinear distribution functions  and the PDF as the same matrix elements of 
gauge-invariant collinear operators, but defined at different scales depending
on the size of the soft momentum involved. Then these functions
are computed with the corresponding 
zero-bin subtractions. Here we also explain why the need for the zero-bin subtraction 
in computing the PDF has remained unnoticed in full QCD and in SCET.
 In Sec.~\ref{jetf}, the jet function, which describes
the collinear particles in the final state in DIS, is computed to one loop including the
zero-bin subtraction. In Sec.~\ref{facone}, all the one-loop results are collected to 
write the kernel $W$. And the renormalization group behavior of 
 the hard function, the kernel and the PDF is discussed. A conclusion
is presented in Sec.~\ref{conc}. In appendix A, the computation of the soft function
at one loop with the nonzero offshellness of the external particles as the IR regulators
is presented, and the collinear quark distribution functions and the quark PDF
in the same regularization method is  presented  in appendix B.
And it is also shown that the IR divergence and the mixed divergence cancel in the  kernel.  
    
\section{Factorization theorems in SCET \label{factscet}} 
In full QCD, the idea of factorization is probed in detail~\cite{Sterman:1986aj} to
separate the hard, collinear and soft parts. In describing the soft part, the eikonal 
approximation is employed to show that it is decoupled from the collinear part.
In SCET, the procedure of the factorization
is elaborated since the factorization is achieved at the operator level. The collinear
and soft parts are expressed in terms of the matrix elements of the operators, and
they do not interact with each other from the outset by introducing the collinear and 
soft fields without any interaction between them \cite{Bauer:2001yt}.

 Factorization theorems involve disparate scales, and it is more transparent and
convenient to employ SCET to see the physics clearly. 
There are two scales $Q$ and $E$ with $Q \gg E$, and the effective
theories can be constructed by integrating out large scales successively, and the
matching between the theories can be performed systematically. The first
effective theory \one\ can be constructed from full QCD
by integrating out the degrees of freedom of order $Q$,  and it describes physics 
with the scale $E < \mu < Q$. The second effective theory \two\ is constructed from \one\ by
integrating out the degrees of freedom of order $E$, and it describes physics with the 
scale $\Lambda_{\mathrm{QCD}} \ll \mu < E$.

The scattering cross sections near the threshold are schematically written  in SCET as
\begin{equation} \label{confac} 
d\sigma =\left\{ \begin{array}{l}
H_{\mathrm{DY}}(Q,\mu) \otimes S_{\mathrm{DY}}(\mu) 
\otimes f_{q/N_1} (\mu)
\otimes f_{\bar{q}/N_2} (\mu),   \\ H_{\mathrm{DIS}}
(Q,\mu)\otimes J(Q\sqrt{1-z},\mu) \otimes S_{\mathrm{DIS}} (\mu) 
\otimes f_{q/N}(\mu),   \\
H_{\mathrm{Higgs}} (Q, \mu) \otimes S_{\mathrm{Higgs}} (\mu) 
\otimes f_{g/N_1} (\mu) \otimes f_{g/N_2} (\mu), 
\end{array} \right. 
\end{equation} 
in DY~\cite{Becher:2007ty}, DIS~\cite{Chay:2005rz,Idilbi:2007ff,Chen:2006vd} and
Higgs~\cite{Ahrens:2008nc} production processes respectively. 
In Drell-Yan processes, and in DIS, the endpoint refers to the hadronic endpoint, and in 
Higgs production it referes to the partonic endpoint. $H(Q,\mu)$ is the hard 
function,  $Q$ is the large scale, and $\mu$ is the 
renormalization scale in the region $E < \mu <Q$.  $E$ is the intermediate scale 
$E\sim Q(1-z) \ll Q$, where $z$ is the Bjorken variable. 
We consider the endpoint region,
$z\rightarrow 1$, but assume $E \gg \Lambda_{\mathrm{QCD}}$ such that perturbation is valid.
$S(\mu)$ is the 
soft function, which is defined in terms of soft Wilson lines. The functions 
$f_{q/N} (\mu)$ and $f_{g/N} (\mu)$ are
the collinear quark and gluon distribution functions defined in terms of the gauge-invariant
collinear operators.  The final-state jet function $J$ appears only in DIS, and describes
the collinear final-state particles. The symbol $\otimes$ means an appropriate
convolution, and the  formulae with the explicit convolution will be presented
below. 

Eq.~(\ref{confac}) is the factorization theorem in \one, based on the idea of the  traditional 
threshold factorization theorem. However, as they stand, the soft
functions $S_{\mathrm{DY}}$, $S_{\mathrm{DIS}}$ and $S_{\mathrm{Higgs}}$ contain
 IR and mixed divergence, so do the collinear functions $f_{q/N_1}$, 
$f_{\bar{q}/N_2}$ and $f_{g/N}$. 
The fact that each factorized part can be computed  separately has a
merit, but the existence of IR divergences  hinders us from using
the evolution of each part. It can be cured by reorganizing the collinear and soft parts
in such a way that the IR divergence is absent except in the PDF. The procedure 
of reorganization is not arbitrary, but is based on
the physical principle that  consistent separation of the collinear and soft modes 
should be maintained at higher loops.  This procedure is accomplished when the scattering
cross sections are written in \two.

It should be noted that not all the soft functions in various processes near the endpoint 
are infrared divergent. For example, the soft function is IR-finite for event shapes in $e^+ e^-$
annihilation \cite{Hornig:2009kv}. And the list of threshold soft functions is given in
Ref.~\cite{Bauer:2010vu}. The IR-finiteness of the soft functions depends on the physical observables
described by the soft functions. In this case, the collinear part is either absent or modified to yield
IR-finite results. Then the soft functions and the collinear functions have physical meaning and 
the evolution of them can be considered. What we deal with here is the case in which the soft function,
and in turn, the collinear distribution functions are IR-divergent. We describe the factorization
procedure when the IR divergence appears in each factorized part.

The collinear distribution functions $f_{q/N}$ and $f_{g/N}$ are the matrix elements of the
collinear operators, and they should be distinguished from the PDF.
In Ref.~\cite{Korchemsky:1992xv}, the PDF has been constructed to include the soft Wilson lines
near the endpoint considering the fact that only real soft gluons can be emitted due to kinematics.
The corresponding PDF in SCET has been considered in Refs.~\cite{Idilbi:2006dg,Becher:2006mr}. 
In the second line of Eq.~(\ref{confac}), $S_{\mathrm{DIS}}(\mu) \otimes f_{q/N} (\mu)$
corresponds
to the PDF near the endpoint in \one, but in Drell-Yan or Higgs production, the combination of the
soft function and the collinear distribution functions do not yield the product of the PDFs.
We reserve the terminology ``parton distribution functions''
 for the same matrix elements 
as those for $f_{q/N}$ and $f_{g/N}$, but evaluated at a much 
lower scale than $E$. 
The PDF will be denoted as $\phi_{q/N}$ and $\phi_{g/N}$ from now on.

When we consider 
DY and DIS processes in \one, there appears a back-to-back current which can be written 
in SCET at leading order as 
\begin{equation} \label{current} 
\overline{q} \gamma^{\mu} q=
C(Q,\mu) \overline{\chi}_{\bar{n}} Y_{\bar{n}}^{\dagger} \gamma_{\perp}^{\mu}
Y_n \chi_n +\mathrm{h.c.}, 
\end{equation}
where $\chi_n = W_n^{\dagger} \xi_n$
is the $n$-collinear  fermion field, and $\chi_{\bar{n}}
=W_{\bar{n}}^{\dagger} \xi_{\bar{n}}$ is the
$\overline{n}$-collinear fermion. The collinear Wilson line $W_n$ 
is introduced to make the current collinear gauge invariant and is given by
\begin{equation}  \label{cowilson}
W_n =\sum_{\mathrm{perm}} \exp \Bigl[
-\frac{g}{\overline{n}\cdot \mathcal{P}+i0} \overline{n}\cdot A_n \Bigr],
\end{equation} 
where $n$ and $\overline{n}$ are lightcone vectors satisfying
$n^2 =\overline{n}^2 =0$, $n\cdot \overline{n}=2$. The operator $\overline{n}\cdot
\mathcal{P}$ extracts the label momentum, and the bracket means the operator 
acts only inside the bracket. The $\overline{n}$-collinear
Wilson line $W_{\bar{n}}$ is obtained by switching $n$ and $\overline{n}$ in 
Eq.~(\ref{cowilson}). In the Higgs production, a back-to-back current with gluons is involved,
and the detailed analysis is given in Sec.~\ref{higgsfac}.

In addition to the UV and IR divergences,  there may exist another type of divergence,
called the rapidity divergence, which occurs when an energetic particle 
emits collinear gluons with infinite rapidity. When all the Feynman diagrams 
are added for a given quantity, the rapidity divergences cancel. But they appear in
individual Feynman diagrams and a regularization is necessary to treat this 
divergence. It is suggested to include rapidity regulators to all the Wilson lines \cite{Chiu:2009yx}. 
A simpler method to regulate the rapidity divergence is to insert 
rapidity regulators only in the collinear Wilson lines as \cite{Chay:2012mh}
\begin{equation} \label{cowilson1}
W_n =\sum_{\mathrm{perm}} \exp \Bigl[
-\frac{g}{\overline{n}\cdot \mathcal{P}+\delta_n+i0} \overline{n}\cdot A_n  \Bigr].
\end{equation} 
The collinear Wilson line $W_n$ can be heuristically derived by attaching $n$-collinear
gluons to all the collinear or heavy particles not in the $n$ direction, and
integrating out the intermediate states. If these particles are on
the mass shell, the collinear Wilson line in Eq.~(\ref{cowilson}) is 
obtained \cite{Chay:2003ju}.  If those particles are slightly offshell, the collinear
Wilson line takes the form in Eq.~(\ref{cowilson1}), where $\delta_n$ can be
related to the combined offshellness of all the collinear and heavy particles not in the $n$ direction.
In this case,  the gauge invariance of the collinear Wilson line is broken. But as long as the 
sum of all the diagrams is gauge-invariant, the introduction of the offshellness 
can be regarded just as an intermediate step.  On the other hand, the regulator 
$\delta_n$ can be regarded simply as a regulator with no relation to the offshellness.
The gauge-invariant formulation with the regulator can be probed, but we use 
Eq.~(\ref{cowilson1}) as it is to regulate the rapidity divergence. We will show 
the dependence on the rapidity regulator and its cancellation, when summed,
at one loop below.

In Eq.~(\ref{current}), the collinear field $\xi_n$
is redefined to be decoupled from the soft interactions as  \cite{Bauer:2001yt}
\begin{equation}
\xi_n \rightarrow Y_n \xi_n, \\ A_n \rightarrow Y_n A_n Y_n^{\dagger},
\end{equation} 
where the soft Wilson line $Y_n$ is given by 
\begin{equation}  \label{softwilson}
Y_n = \sum_{\mathrm{perm}} \exp \Bigl[ -\frac{g}{n\cdot \mathcal{R} +i0} n\cdot
A_s\Bigr]. 
\end{equation} 
Here $\mathcal{R}^{\mu} =i\partial^{\mu}$ is a derivative operator on the soft field, 
which extracts soft momentum.  In Eq.~(\ref{softwilson}), 
the soft Wilson lines are  collectively written as $Y_n$, but they should be determined  
according to the kinematic situations of the scattering process. The detailed prescription for 
the soft Wilson lines  is discussed in Ref.~\cite{Chay:2004zn}, and it will be used in defining 
the soft  function.

The matching between QCD and
\one\ is performed by comparing the matrix elements
of the same quantities at the scale $Q$. The Wilson coefficient $C(Q^2,\mu)$ in 
Eq.~(\ref{current}), for example, is the matching coefficient for the back-to-back
current. Since the IR divergence
is cancelled in the matching, the Wilson coefficients are free of IR divergence. 
Now we go further down below $E$, and the degrees of freedom above $E$ are
integrated out to produce the final effective theory,
\two. The matching between \one\ and \two\ can also be performed systematically.

The key ingredient in adopting the successive effective theories is to relate the
collinear quark and gluon distribution functions $f_{q/N}$ and $f_{g/N}$ in \one\ to the 
quark and gluon PDF $\phi_{q/N}$ and $\phi_{g/N}$ in \two\ as \cite{Chay:2012jr}
\begin{eqnarray} \label{injet} 
f_{q/N} (x,\mu) &=&\int_x^1 \frac{dz}{z} K_{qq} (z,\mu) \phi_{q/N} (x/z,\mu),
\nonumber \\
f_{g/N} (x,\mu) &=&\int_x^1 \frac{dz}{z} K_{gg} (z,\mu) \phi_{g/N} (x/z,\mu). 
\end{eqnarray}
We distinguish the collinear distribution functions $f_{q/N}$ and $f_{g/N}$ 
from the standard  PDF $\phi_{q/N}$ and $\phi_{g/N}$ though the operator 
definitions for both are the same, but 
because they are evaluated at different energy scales. The collinear distribution
functions are evaluated at an intermediate energy scale $E$
($\Lambda_{\mathrm{QCD}} \ll E \ll Q$), while the PDF are evaluated at the scale
much below $E$, but much larger than $\Lambda_{\mathrm{QCD}}$. 
As will be shown below, this distinction is important. We call $K_{qq}$ and $K_{gg}$ 
the initial-state quark and gluon jet functions respectively. 
The initial-state jet function looks like the matching coefficient 
for the collinear matrix elements between $\mathrm{SCET}_{\mathrm{I}}$ and
$\mathrm{SCET}_{\mathrm{II}}$, but strictly speaking it is not. It includes
IR divergences unlike the Wilson coefficient $C(Q^2,\mu)$. 
 The initial-state jet function is actually the difference
in the contributions of the soft modes in the two effective theories.

The ``initial-state jet function'' should not be confused since the same terminology was also 
employed 
independently in  Ref.~\cite{Stewart:2009yx}, but the physical meaning and the IR structure are different.  The initial-state jet function in the beam function is an exclusive quantity
which involves an observable parameter acting as an IR cutoff, hence IR-finite.
On the other hand, the initial-state jet function here
 results from the inclusive scattering
cross section. Since there is no IR regulator in our case, special care is needed to handle the IR
divergence.

The soft function in \one\ describes the contributions of the soft momentum of order 
$E\sim Q(1-z)$, hence the loop momentum of order $E\sim Q(1-z)$ should be subtracted
from the collinear part to avoid double counting through the zero-bin subtraction.
In \two, the corresponding zero-bin subtraction should be performed with the momentum much
smaller than $Q(1-z)$. This small momentum is sometimes referred to as the 
ultrasoft (usoft) momentum. These
two kinds of zero-bin subtractions are different, and their difference yields 
the initial-state jet function. If there is no distinction between \one\
and \two,  for example, away from the endpoint, 
the collinear distribution function and the PDF are identical and 
$K_{qq}(z,\mu)= K_{gg} (z,\mu) =\delta (1-z)$ to all orders in $\alpha_s$, and there 
are no soft functions.

The initial-state jet function contains IR divergence, so does the soft function. 
However, the sum of these two functions is free of IR divergence. Furthermore
the mixing of the UV and IR divergences is removed only in this sum. 
Therefore only the sum is physically meaningful and constitutes a component in the factorization
theorem.  Instead, we would like to obtain the factorization theorems, in which each part
is IR finite. What we propose as the proper factorization theorem is to use Eq.~(\ref{injet})
to express the  factorization theorem from \one\ to the factorization theorem in \two\ as
\begin{eqnarray} \label{scetfac} 
d\sigma _{\mathrm{DY}} &=& H_{\mathrm{DY}}
(Q,\mu) \otimes S_{\mathrm{DY}} (E,\mu) \otimes K_{qq} (E,\mu) \otimes
K_{\bar{q}\bar{q}} (E,\mu) \otimes \phi_{q/N_1} (\mu)\otimes
\phi_{\bar{q}/N_2}(\mu) \nonumber \\ &=& H_{\mathrm{DY}} (Q,\mu) \otimes
W_{\mathrm{DY}} (E,\mu) \otimes \phi_{q/N_1} (\mu)\otimes
\phi_{\bar{q}/N_2}(\mu), \nonumber \\ 
d\sigma_{\mathrm{DIS}} &=&
H_{\mathrm{DIS}} (Q,\mu)\otimes J(Q\sqrt{1-z},\mu) \otimes S_{\mathrm{DIS}}
(E,\mu) \otimes K_{qq} (E,\mu) \otimes \phi_{q/N} (\mu)\nonumber \\ &=&
H_{\mathrm{DIS}} (Q,\mu)\otimes J(Q\sqrt{1-z},\mu) \otimes W_{\mathrm{DIS}}
(E,\mu) \otimes \phi_{q/N} (\mu),  \nonumber \\
d\sigma _{\mathrm{Higgs}} &=& H_{\mathrm{Higgs}}
(Q,\mu) \otimes S_{\mathrm{Higgs}} (E,\mu) \otimes K_{gg} (E,\mu) \otimes
K_{gg} (E,\mu) \otimes \phi_{g/N_1} (\mu)\otimes
\phi_{g/N_2}(\mu) \nonumber \\ &=& H_{\mathrm{Higgs}} (Q,\mu) \otimes
W_{\mathrm{Higgs}} (E,\mu) \otimes \phi_{g/N_1} (\mu)\otimes
\phi_{g/N_2}(\mu),
\end{eqnarray} 
for DY, DIS and the Higgs production processes respectively. 
Here we define the soft kernels $W$ as 
\begin{eqnarray} \label{kernelw}
W_{\mathrm{DY}} (E,\mu) &=& S_{\mathrm{DY}} (E,\mu) \otimes K_{qq} (E,\mu)
\otimes K_{\bar{q}\bar{q}} (E,\mu), \nonumber \\ 
W_{\mathrm{DIS}} (E,\mu) &=&S_{\mathrm{DIS}} (E,\mu) \otimes K_{qq} (E,\mu),
\nonumber \\
W_{\mathrm{Higgs}} (E,\mu) &=& S_{\mathrm{Higgs}} (E,\mu) \otimes K_{gg} (E,\mu)
\otimes K_{gg} (E,\mu).
\end{eqnarray} 
Though the initial-state jet function is included in the kernel $W$, we call this the
soft kernel since the initial-state jet function consists of the soft limits of the
collinear parts, as will be explained below. Eq.~(\ref{scetfac}) is our new 
factorization theorem, in which each factorized part can be systematically
computed in perturbation theory. Furthermore, all the factors except the PDF are
IR finite \cite{Chay:2012jr}. The explicit convoluted form and the one-loop results are shown 
in this paper.

One may wonder why we go through this labyrinthine procedure, while we know that
the PDF in full QCD can be obtained by a straightforward calculation without the 
problem of IR divergence. It will be shown explicitly how full QCD works later. But
to put it simply, the answer lies in the fact that full QCD computations correspond to
the calculation of the PDF in \two. In \two, the usoft zero-bin contribution simply vanishes,
hence no effect results from the usoft zero-bin subtraction.
The details will be explained in Sec~\ref{colli}. 

\subsection{Drell-Yan process near the endpoint\label{dyfac}} 
Let us consider the inclusive Drell-Yan process
$p (P_1) \overline{p} (P_2) \rightarrow l^+ l^- (q) + X (p_X)$, where $l^+ l^-$
are a lepton pair and $X$ denotes hadrons in the final state. We define the
structure function $F_{\mathrm{DY}} (\tau)$ as 
\begin{equation} \label{fdy}
F_{\mathrm{DY}} (\tau) = -N_c \int \frac{d^4 q}{(2\pi)^4} \theta (q^0) \delta
(q^2 -s \tau) \int d^4 z e^{-i q\cdot z} \langle N_1 N_2 | J_{\mu}^{\dagger} (z)
J^{\mu} (0)|N_1 N_2 \rangle, 
\end{equation} 
with the number of colors $N_c$, and $\tau = Q^2 /s$, where $Q^2$ is 
the invariant mass squared of the lepton pair, and
$s$ is the hadronic center-of-mass energy squared. Near the
endpoint $\tau \rightarrow 1$,  the final-state particles are either soft with
the interaction, or $n$- and $\overline{n}$-collinear without the interaction. The
differential scattering cross section is given as $d\sigma/d\tau = \sigma_0
F_{\mathrm{DY}} (\tau)$, where $\sigma_0 = 4\pi \alpha^2 Q_f^2/(3N_c Q^2)$ is
the Born cross section for the quark flavor $f$ with the electric charge $Q_f$.

Now we express $F_{\mathrm{DY}} (\tau)$ in SCET by using the current in
Eq.~(\ref{current}), and the final state $|X\rangle$ is decomposed as $|X\rangle
=|X_n\rangle |X_{\bar{n}}\rangle |X_s\rangle$, the $n$-, 
$\overline{n}$-collinear and the soft states. The momentum $q^{\mu}$ of
the lepton pair is given by $q= P_1 +P_2 -p_X$, where $P_{1,2}$ is the momentum
of the hadron $N_{1,2}$ in the $n$ and $\overline{n}$ directions respectively.
These momenta are given by $P_1^{\mu} = \overline{n} \cdot P_1 n^{\mu}/2$,
and $P_2^{\mu} = n\cdot P_2 \overline{n}^{\mu}/2$ where  
 $s =\overline{n}\cdot P_1 n\cdot P_2$. 
The momentum of the final-state particles $p_X$ can also be
decomposed as $p_X = p_{X_n} +p_{X_{\bar{n}}} +p_{X_s}$, and $q$ can be written
as 
\begin{equation}  \label{traq}
q=P_1 +P_2 -( p_{X_n} +p_{X_{\bar{n}}} +p_{X_s}) = (P_1
-p_{X_n}) + (P_2 -p_{X_{\bar{n}}}) -p_{X_s} = p_1 +p_2 -p_{X_s}, 
\end{equation}
where $p_{1,2}$ are the momenta of the incoming partons inside the hadrons
$N_{1,2}$ respectively. From now on, we express Eq.~(\ref{fdy}) in terms of the
partonic variables. First the argument in the delta function in the partonic center-of-mass
frame can be written as
\begin{eqnarray} 
q^2 -s\tau &=& q^2 -Q^2 = (p_1 +p_2)^2 -2p_{X_s} \cdot (p_1
+p_2) -Q^2 \nonumber \\ 
&=& \hat{s} -2\eta \hat{s}^{1/2} -Q^2 \sim \hat{s}
\Bigl(1-z-\frac{2\eta}{Q}\Bigr), 
\end{eqnarray} 
where $\eta = v\cdot p_{X_s} = (n\cdot p_{X_s} +\overline{n}\cdot
p_{X_s})/2 = p_{X_s}^0$, and $z=Q^2/\hat{s}$, with the center-of-mass energy 
squared $\hat{s}$ for the partons. We neglected $p_{X_s}^2$
in the first line. Near the endpoint, $\tau <z <1$, 
$\tau \rightarrow 1$, and higher powers of $1-z$ are neglected.

The structure function can be written in SCET as 
\begin{eqnarray}\label{dyfactor} 
F_{\mathrm{DY}} (\tau) &=& -\frac{N_c}{\hat{s}} H_{DY} (Q,\mu)
 \langle N_1 N_2| \overline{\chi}_n Y_n^{\dagger}
\gamma_{\mu}^{\perp} Y_{\bar{n}} \chi_{\bar{n}} \overline{\chi}_{\bar{n}} \delta
\Bigl(1-z+\frac{2v\cdot \mathcal{R}}{Q}\Bigr) Y_{\bar{n}}^{\dagger} Y_n
\gamma_{\mu}^{\perp} \chi_n |N_1 N_2\rangle \nonumber \\ 
&=& -N_c  H_{DY}
(Q,\mu)  \int \frac{dy_1 dy_2}{\hat{s}} \langle N_1 N_2 | \overline{\chi}_n
Y_n^{\dagger} \gamma_{\mu}^{\perp} Y_{\bar{n}} \chi_{\bar{n}} \nonumber \\
&&\times \overline{\chi}_{\bar{n}} \delta \Bigl(y_2 
+\frac{n\cdot \mathcal{P}^{\dagger}}{n\cdot
P_2} \Bigr) \delta \Bigl(1-z+\frac{2v\cdot \mathcal{R}}{Q}\Bigr)
Y_{\bar{n}}^{\dagger} Y_n \gamma_{\mu}^{\perp} \delta \Bigl( y_1
-\frac{\overline{n}\cdot \mathcal{P}}{\overline{n}\cdot P_1} \Bigr) \chi_n |N_1
N_2\rangle, 
\end{eqnarray} 
where $H_{DY} (Q^\mu) =|C_{DY}(Q,\mu)|^2$ is the
hard function and $\hat{s} =y_1 y_2 s$. From the last expression 
in Eq.~(\ref{dyfactor}), the soft interactions
are decoupled, and the $n$- and $\overline{n}$-collinear parts are also
decoupled since they no longer communicate to each other in SCET. The collinear
matrix elements, after the operators are Fierz transformed and decoupled, can be
written as 
\begin{eqnarray} \label{codi} 
\langle N_1 | \Bigl[ \delta \Bigl( y_1
-\frac{\overline{n}\cdot \mathcal{P}}{\overline{n}\cdot P_1} \Bigr)
\chi_n\Bigr]_a^{\alpha} \Bigl[\overline{\chi}_n \Bigr]_b^{\beta} |N_1\rangle &=&
\frac{\overline{n}\cdot P_1}{2N_c} \delta^{\alpha \beta} 
\Bigl(\frac{\FMslash{n}}{2} \Bigr)_{ab} f_{q/N_1} (y_1), \nonumber \\ \langle
N_2 |\Bigl[\chi_{\bar{n}} \Bigr]_{a}^{\alpha} \Bigl[ \overline{\chi}_{\bar{n}}
\delta \Bigl(y_2 +\frac{n\cdot \mathcal{P}^{\dagger}}{n\cdot P_2}\Bigr)
\Bigr]_b^{\beta} |N_2 \rangle &=& \frac{n\cdot P_2}{2N_c} \delta^{\alpha\beta} 
\Bigl(\frac{\FMslash{\overline{n}}}{2}\Bigr)_{ab} f_{\bar{q}/N_2} (y_2),
\end{eqnarray} 
where $\alpha$, $\beta$ are color indices, and $a$, $b$ are Dirac
indices. Eq.~(\ref{codi}) defines the collinear quark and antiquark distribution functions 
for the incoming partons as 
\begin{eqnarray} f_{q/N_1} (x_1) &=& \langle N_1|
\overline{\chi}_n \frac{\FMslash{\overline{n}}}{2} \delta (x_1 \overline{n}\cdot
P_1 -\overline{n} \cdot \mathcal{P}) \chi_n |N_1\rangle, \nonumber \\
f_{\bar{q}/N_2}(x_2) &=& \langle N_2| \overline{\chi}_{\bar{n}}
\frac{\FMslash{n}}{2} \delta (x_2 n\cdot P_2 +n\cdot \mathcal{P}^{\dagger})
 \chi_{\bar{n}}|N_2\rangle. 
\end{eqnarray}
These collinear distribution functions are defined in \one, that is, the matrix elements are 
evaluated at the scale above $E=Q(1-z)$.

Finally the factorized form for $F_{DY} (\tau)$ is written as 
\begin{eqnarray} \label{convfdy}
F_{DY} (\tau) &=& H_{DY} (Q,\mu) \int \frac{dy_1}{y_1} \frac{dy_2}{y_2}
f_{q/N_1} (y_1) f_{\bar{q}/N_2} (y_2) S_{\mathrm{DY}} (z,\mu) \nonumber \\ 
&=&H_{DY} (Q,\mu) \int_{\tau}^1 \frac{dz}{z} 
S_{\mathrm{DY}} (z,\mu) F_{q\bar{q}} \Bigl( \frac{\tau}{z}\Bigr), 
\end{eqnarray}
where the soft function $S_{\mathrm{DY}}(z,\mu)$ is defined as 
\begin{equation} 
S_{\mathrm{DY}}  (z,\mu) = \frac{1}{N_c}
\langle 0|\mathrm{tr} Y_n^{\dagger} Y_{\bar{n}} \delta \Bigl( 1-z+\frac{2
v\cdot \mathcal{R}}{Q}\Bigr) Y_{\bar{n}}^{\dagger} Y_n|0\rangle, 
\end{equation}
and is normalized to $\delta (1-z)$ at tree level. The soft Wilson lines 
$Y_{\bar{n}}^{\dagger}$ and $Y_n$ are chosen such that the antiquark and the quark 
come from $-\infty$ \cite{Chay:2004zn}, and their hermitian conjugates are employed
in the left-hand side of the delta function. The function $F_{q\bar{q}}$ is given as 
\begin{equation} 
F_{q\bar{q}}
\Bigl(\frac{\tau}{z}\Bigr) = \int_{\tau/z}^1 \frac{dy_1}{y_1} f_{q/N_1} (y_1)
f_{\bar{q}/N_2} \Bigl(\frac{\tau}{zy_1}\Bigr). 
\end{equation}
Eq.~(\ref{convfdy}) is the result of the traditional threshold factorization. 

In our factorization scheme, we further express the collinear function in terms of the PDF as 
\begin{eqnarray}\label{fphi} 
f_{q/N} (x,\mu) &=&\int_x^1 \frac{dz}{z} K_{qq} (z,\mu) \phi_{q/N}
(x/z,\mu), \nonumber \\ 
f_{\bar{q}/N} (x,\mu) &=&\int_x^1 \frac{dz}{z}
K_{\bar{q}\bar{q}} (z,\mu) \phi_{\bar{q}/N} (x/z,\mu). 
\end{eqnarray}
By shuffling the order of integrations, and changing the integration variables,
$F_{\mathrm{DY}}$ is written as 
\begin{eqnarray}  \label{fdyour}
F_{\mathrm{DY}} (\tau)&=&
H_{\mathrm{DY}}(Q,\mu) \int_{\tau}^1 \frac{dw}{w} \int_{\tau/w}^1 \frac{dx}{x}
\phi_{q/N_1} (x,\mu) \phi_{\bar{q}/N_2} \Bigl(\frac{\tau}{xw},\mu\Bigr)
\nonumber \\ 
&\times&\int_w^1 \frac{dz}{z}  S_{\mathrm{DY}} (z,\mu) \int_{w/z}^1
\frac{dt}{t} K_{qq} (t,\mu) K_{\bar{q}\bar{q}} \Bigl(\frac{w}{zt},\mu\Bigr)
\nonumber \\ 
&=& H_{\mathrm{DY}}(Q,\mu) \int_{\tau}^1 \frac{dw}{w}
W_{\mathrm{DY}} (w,\mu ) \int_{\tau/w}^1 \frac{dx}{x} \phi_{q/N_1} (x,\mu)
\phi_{\bar{q}/N_2} \Bigl(\frac{\tau}{xw},\mu\Bigr) \nonumber \\ 
&=&
H_{\mathrm{DY}}(Q,\mu) \int_{\tau}^1 \frac{dw}{w} W_{\mathrm{DY}} (w,\mu ) 
\Phi_{q\bar{q}} \Bigl(\frac{\tau}{w},\mu \Bigr), 
\end{eqnarray} 
where $W_{\mathrm{DY}}$ and $\Phi_{q\bar{q}}$ are given as 
\begin{eqnarray}  \label{defwdy}
W_{\mathrm{DY}} (w,\mu) &=& \int_w^1
\frac{dz}{z}  S_{\mathrm{DY}} (z,\mu) \int_{w/z}^1 \frac{dt}{t} K_{qq} (t,\mu)
K_{\bar{q}\bar{q}} \Bigl(\frac{w}{zt},\mu\Bigr), \nonumber \\ 
\Phi_{q\bar{q}}
\Bigl(\frac{\tau}{w},\mu \Bigr) &=& \int_{\tau/w}^1\frac{dx}{x} \phi_{q/N_1}
(x,\mu) \phi_{\bar{q}/N_2} \Bigl(\frac{\tau}{xw},\mu\Bigr). 
\end{eqnarray}
Eq.~(\ref{fdyour}) is our new factorization theorem for DY process near the endpoint.
 The structure function is written as the convolution of the hard function, the soft 
kernel $W_{\mathrm{DY}}$, and the product of the PDFs.
And it will be shown that $W_{\mathrm{DY}}$ is IR finite at one loop.
 
\subsection{Deep inelastic scattering near the endpoint\label{disfac}}

Consider the inclusive deep inelastic scattering $e (k_1) +p (P) \rightarrow e
(k_2) + X (p_X)$ near the endpoint. The hadronic tensor $W^{\mu\nu}$ is written
as 
\begin{eqnarray} 
W^{\mu\nu} &=& \frac{1}{2\pi} \sum_X \int d^4 z e^{iq\cdot
z} \langle N(P)|J^{\mu \dagger} (z) |X (p_X)\rangle \langle X (p_X) |J^{\nu}
(0)|N(P)\rangle \\ 
&=& \frac{1}{2\pi} \sum_X (2\pi)^4 \delta^{(4)} (q+P-p_X)
\langle N|J^{\mu \dagger} |X\rangle \langle X|J^{\nu}|N\rangle, \nonumber
\end{eqnarray}
 where $q=k_2-k_1$ is the momentum transfer from the leptonic
system. The summation $\sum_X$ includes all the phase spaces of the final-state
particles. Here we consider the electromagnetic current only, but the inclusion of the weak
current is straightforward. In the Breit frame, the incoming proton is in the $n$ direction, and
the outgoing particles are in the $\overline{n}$ direction. The analysis near the endpoint in DIS
using SCET was first performed both in the Breit and the target-rest frames in Ref.~\cite{Manohar:2003vb}.

The momentum $q$ and the momentum $P$ of the incoming proton  can be written as 
\begin{equation} 
q^{\mu}= \frac{\overline{n}^{\mu}}{2} Q -\frac{n^{\mu}}{2}Q, \ P^{\mu}
=\overline{n}\cdot P \frac{n^{\mu}}{2}, 
\end{equation} 
and the Bjorken variable is defined as 
\begin{equation} 
x=\frac{Q^2}{2P\cdot q} \sim
\frac{Q}{\overline{n}\cdot P}.
\end{equation}
The endpoint corresponds to the limit $x\rightarrow 1$. 

The hadronic tensor can be expressed in terms of the structure functions as \cite{Chay:2005rz}
\begin{equation} 
W^{\mu\nu} = -g_{\perp}^{\mu\nu} F_1 +v^{\mu} v^{\nu} F_L,
\end{equation}
with $v^{\mu} = (n^{\mu} +\overline{n}^{\mu})/2$. The structure
function $F_1$ is given by 
\begin{eqnarray} 
F_1 &=& -\frac{1}{2}
g_{\perp}^{\mu\nu} W_{\mu\nu} =-\frac{1}{4\pi} \sum_X (2\pi)^4 \delta^{(4)}
(q+P-p_X) \langle N|J^{\mu \dagger} |X\rangle \langle X|J_{\mu}|N\rangle \\ 
&=&
-\frac{1}{4\pi} \sum_X (2\pi)^4 \delta^{(4)} (q+P-p_X) 
|C_{\mathrm{DIS}} (Q,\mu)|^2 \nonumber \\
&&\times \int_0^1 dy \langle N|\overline{\chi}_n Y_n^{\dagger}\gamma^{\mu}_{\perp}
\tilde{Y}_{\bar{n}} \chi_{\bar{n}} |X\rangle \langle X| \overline{\chi}_{\bar{n}}
\tilde{Y}_{\bar{n}}^{\dagger} \gamma_{\mu}^{\perp} Y_n \delta \Bigl(y-\frac{\overline{n}\cdot
\mathcal{P}}{\overline{n}\cdot P}\Bigr) \chi_n |N\rangle, \nonumber
\end{eqnarray}
where the current $J^{\mu}$ in SCET is given by 
\begin{equation} 
J^{\mu}= C_{\mathrm{DIS}} (Q,\mu) \overline{\chi}_{\bar{n}}
\tilde{Y}_{\bar{n}}^{\dagger} \gamma^{\mu}_{\perp}  Y_n \chi_n. 
\end{equation}
Here the soft Wilson lines are chosen such that the incoming quark comes from $-\infty$,
and the outgoing quark goes to $\infty$.

We now define the two collinear functions in the $\overline{n}$ and $n$ directions,
which do not communicate to each other. The final-state jet function in the 
$\overline{n}$ direction is defined as
\begin{equation} 
\sum_{X_{\bar{n}}} \chi_{\bar{n}}
|X_{\bar{n}}\rangle \langle X_{\bar{n}}| \overline{\chi}_{\bar{n}}
=\frac{\FMslash{\overline{n}}}{2} \int \frac{d^4 p_{X_{\bar{n}}}}{(2\pi)^4}
\overline{J}(p_{X_{\bar{n}}}).
\end{equation}
The jet function is a function of $n\cdot p_{X_{\bar{n}}}$ only.
The collinear distribution function in the $n$ direction is defined, 
as in Eq.~(\ref{codi}), by 
\begin{equation} \sum_{X_n} \langle N| (\overline{\chi}_n)^{\alpha}_a
|X_n\rangle \langle X_n| \delta \Bigl( y-\frac{\overline{n}\cdot
\mathcal{P}}{\overline{n}\cdot P} \Bigr) (\chi_n)^{\beta}_b|N\rangle =
\frac{\overline{n}\cdot P}{2N_c} \delta^{\alpha \beta}
\Bigl(\frac{\FMslash{n}}{2} \Bigr)_{ba} f_{q/N} (y).
\end{equation}
In terms of these two collinear functions,
the structure function $F_1$ can be written as 
\begin{eqnarray} \label{f1dis}
F_1(x) &=& |C_{\mathrm{DIS}} (Q,\mu)|^2 \int_x^1 
\frac{dy}{y} f_{q/N} (y,\mu) \nonumber \\
&&\times\int_0^{Q(1-x/y)} \frac{d\overline{n}\cdot k}{2\pi} \overline{J}(
\overline{n}\cdot k,\mu) \overline{S}_{\mathrm{DIS}}\Bigl( 1-\frac{x}{y}
-\frac{\overline{n}\cdot k}{Q}, \mu\Bigr), 
\end{eqnarray} 
where the soft function $\overline{S}_{\mathrm{DIS}}$ is defined as 
\begin{equation}
\overline{S}_{\mathrm{DIS}} (1-z)= \frac{1}{N_c}\langle 0| \mathrm{tr} \Bigl[
Y_n^{\dagger} \tilde{Y}_{\bar{n}} \delta \Bigl( 1-z+\frac{\overline{n}\cdot
\mathcal{R}}{Q}\Bigr) \tilde{Y}_{\bar{n}}^{\dagger} Y_n \Bigr] |0\rangle.
\end{equation}
Changing the variables $\overline{n}\cdot k = (1-w)Q$, and defining 
\begin{equation} \label{js} 
J(w,\mu) =\frac{Q}{2\pi} \overline{J} \Bigl(Q(1-w),
\mu\Bigr), \ \ S_{\mathrm{DIS}}\Bigl(\frac{z}{w},\mu\Bigr) = w
\overline{S}_{\mathrm{DIS}} (w-z), 
\end{equation}
The jet function $J(w,\mu)$
and the soft function $S_{\mathrm{DIS}}(w,\mu)$ are normalized to $\delta (1-w)$
respectively at tree level.
Eq.~(\ref{f1dis}) is written as 
\begin{equation} \label{convf1} 
F_1 (x) = H_{\mathrm{DIS}} (Q,\mu) \int_x^1
\frac{dz}{z} f_{q/N} \Bigl(\frac{x}{z},\mu\Bigr) \int_z^1 \frac{dw}{w} J(w,\mu)
S_{\mathrm{DIS}} \Bigl(\frac{z}{w},\mu\Bigr), 
\end{equation}
where
$H_{\mathrm{DIS}} (Q,\mu) =|C_{\mathrm{DIS}}(Q,\mu)|^2$, and subleading terms of
$S_{\mathrm{DIS}}$ in $w$ is neglected near the endpoint. 

Eq.~(\ref{convf1}) is the result of the traditional factorization theorem near
the endpoint. In our formulation, as was done in DY process, we go further by
invoking Eq.~(\ref{fphi}) to write Eq.~(\ref{convf1}) as 
\begin{equation} 
F_1 (x) = H_{\mathrm{DIS}} (Q,\mu)\int_x^1 \frac{dz}{z} \int_{x/z}^1 \frac{dy}{y} 
K_{qq} (y) \phi \Bigl( \frac{x}{yz} \Bigr) \int_z^1 \frac{dw}{w} J(w) 
S_{\mathrm{DIS}}\Bigl( \frac{z}{w}\Bigr).
\end{equation}
Changing the order of integrations successively, and redefining the integration variables, 
the factorized structure function is given as
\begin{eqnarray} \label{f1our}
F_1(x) &=& H(Q,\mu) \int_x^1 \frac{dw}{w} J(w) \int_{x/w}^1 \frac{dv}{v} 
\Bigl[ \int_v^1 \frac{du}{u}  S_{\mathrm{DIS}} (u, \mu) K_{qq}
\Bigl( \frac{v}{u},\mu\Bigr) \Bigr] 
\phi_{q/N} \Bigl(\frac{x}{vw},\mu\Bigr) \nonumber \\
&=& H(Q,\mu) \int_x^1 \frac{dw}{w} J(w) \int_{x/w}^1 \frac{dv}{v} 
W_{\mathrm{DIS}} (v,\mu) \phi_{q/N} \Bigl(\frac{x}{vw},\mu\Bigr),
\end{eqnarray}
where $W_{\mathrm{DIS}} (v,\mu)$ is given by
\begin{equation} \label{defwdis}
W_{\mathrm{DIS}} (v,\mu) = \int_v^1 \frac{du}{u}  S_{\mathrm{DIS}} (u, \mu) K_{qq}
\Bigl( \frac{v}{u},\mu\Bigr).
\end{equation}
Eq.~(\ref{f1our}) is our result of the factorization theorem in DIS with successive matching.
The soft kernel $W_{\mathrm{DIS}}$ is IR finite, and surprisingly enough
its radiative corrections to all orders vanish and $W_{\mathrm{DIS}}(v)=
\delta (1-v)$. The argument about why it is true to all orders in $\alpha_s$ 
is given after the one-loop corrections are presented.

\subsection{Higgs production near the partonic threshold \label{higgsfac}}

The factorization proof for the Higgs production is similar to that for DY process except that
the initial-state particles are gluons instead of a $q\overline{q}$ pair. Therefore this 
process involves the gluon PDF and the initial-state gluon jet function.  The effective Lagrangian 
for the Higgs production after integrating out the top quark loop is given by
\begin{equation}
\mathcal{L}_{\mathrm{eff}} = C_t (m_t, \mu) \frac{H}{v} G_{\mu\nu}^a G_{\mu\nu a},
\end{equation}
where $v$ is the electroweak vacuum expectation value, $m_t$ is the top quark mass, 
$H$ is the Higgs field and  
$G_{\mu\nu}^a$ is the field strength tensor for gluons. The coefficient $C_t (m_t^2, \mu)$
to second order in $\alpha_s$, and in the heavy top quark mass limit, 
is given by~\cite{Chetyrkin:1997iv,Kramer:1996iq}
\begin{equation}
C_t (m_t,\mu) = \frac{\alpha_s (\mu)}{12\pi} \Bigl( 1+
\frac{\alpha_s (\mu)}{\pi} \frac{11}{4}\Bigr).
\end{equation}

The gluon field strength tensor is matched onto SCET as
\begin{equation}
G_{\mu\nu}^a G^{\mu\nu a} \rightarrow -2C_H (Q,\mu)
 \mathcal{B}_{\bar{n}}^{\perp\mu}
\mathcal{Y}_{\bar{n}}^{\dagger} \mathcal{Y}_n \mathcal{B}_{n\perp}^{\mu}
=-2C_H (Q,\mu) O_g,
\end{equation}
after decoupling the soft interaction. In this process, $Q=m_H$, the Higgs mass. Here $\mathcal{B}_n^{\perp\mu}$ is defined as
 \begin{equation}
\mathcal{B}_n^{\perp\mu} =\frac{1}{g} [\overline{n}\cdot \mathcal{P} W_n^{\dagger} 
iD_n^{\perp\mu} W_n] =i\overline{n}_{\alpha} g^{\mu}_{\perp\beta} W_n^{\dagger} 
G_n^{\alpha\beta} W_n = i\overline{n}_{\alpha} g^{\mu}_{\perp\beta} T^a
 (\mathcal{W}_n^{\dagger})^{ab} G_n^{\alpha\beta,b}.
\end{equation}
In the final expression, the collinear Wilson line $\mathcal{W}_n$ is 
written in the adjoint representation rather than in the fundamental representation. 
That means that the generator $T^a$ is given by $(T^a)^{bc} = -if^{abc}$. The collinear
Wilson lines can be expressed either in the fundamental representation or in the adjoint representation.
Both approaches are equivalent, but the adjoint representation is employed here since it makes 
the expression in  the factorization formula look similar to that in DY processes. The Wilson coefficient
 $C_H$ is the matching coefficient between the full theory and SCET and is given to order 
$\alpha_s$ by~\cite{Ahrens:2008qu}
\begin{equation} \label{ch}
C_H (Q,\mu) =1+\frac{\alpha_s C_A}{4\pi} \Bigl(-\ln^2 \frac{\mu^2}{Q^2} 
+\frac{7}{6} \pi^2-2i\pi \ln \frac{\mu^2}{Q^2}\Bigr), 
\end{equation}
where $C_A =N_c$. Then the effective Lagrangian in  SCET is given by
\begin{equation}
\mathcal{L}_{\mathrm{SCET}} = -\frac{2}{v} C_t (m_t,\mu) C_H (Q,\mu) HO_g 
= -C(\mu) H O_g.
\end{equation}

The kinematics of the Higgs production process $p (P_1) p(P_2) \rightarrow HX$ is 
similar to that in DY process.  The momentum $q$ of the Higgs particle is given by
\begin{equation}
q=P_1 + P_2 -p_X = p_1 + p_2 -p_{X_s},
\end{equation}
as in Eq.~(\ref{traq}).  The cross section for the Higgs production process is written as 
\begin{eqnarray}
\sigma (pp\rightarrow HX) &=& \frac{\pi}{s} \sum_X \int  d^4 q 
  \delta^{(4)} (P_1 +P_2 -q-p_X) \delta (q^2 -Q^2) \theta (q^0) \nonumber \\
&&\times |C(\mu)|^2 \langle N_1 N_2|O_g^{\dagger} |X\rangle
\langle X| O_g |N_1 N_2\rangle \nonumber \\
&=& \frac{\pi}{s} |C(\mu)|^2 \int dx_1 dx_2
\langle N_1 N_2| \mathcal{B}_n^{\perp\mu a} \cy_n^{\dagger a b} \cy_{\bar{n}}^{bc} 
\mathcal{B}_{\bar{n} \mu}^{\perp c} 
\frac{1}{\hat{s}} \delta \Bigl( 1-z-\frac{2\eta}{Q}\Bigr)  \nonumber \\
&&\times \Bigl[\delta \Bigl(x_2 -\frac{n\cdot \mathcal{P}}{n\cdot P_2}\Bigr) 
\mathcal{B}_{\bar{n}}^{\perp \nu d} \Bigr] \cy_{\bar{n}}^{\dagger d e} \cy_n^{ef}
\Bigl[\delta \Bigl( x_1 -\frac{\overline{n}\cdot \mathcal{P}}{\overline{n}\cdot P_1}\Bigr)
\mathcal{B}_n^{\perp\nu f}\Bigr] |N_1 N_2\rangle,
\end{eqnarray}
where $\eta = v\cdot p_X =p_X^0$ and $z=Q^2/\hat{s}$  ($\tau = Q^2/s$ )
with the center-of-mass energy squared for the partons (the hadrons). 
The partonic threshold corresponds to the limit 
$z\rightarrow 1$.  As in DY process, the soft interactions are decoupled,
so are the $n$- and $\overline{n}$-collinear parts. The collinear matrix elements are written as
\begin{equation} \label{cogf1}
\langle N_1| \mathcal{B}_n^{\perp\mu a} \Bigl[\delta \Bigl( x_1 -
\frac{\overline{n}\cdot \mathcal{P}}{\overline{n}\cdot P_1}\Bigr)
\mathcal{B}_n^{\perp\nu f}\Bigr] |N_1\rangle =  
\frac{g_{\perp}^{\mu\nu} \delta ^{af}}{2(N_c^2 -1)} x(\overline{n}\cdot P_1)^2 f_{g/N_1} (x).  
\end{equation}
Inverting  Eq.~(\ref{cogf1}), the collinear gluon distribution amplitude $f_{g/N_1}$ is
written as
\begin{equation} \label{defcg}
 f_{g/N_1} (x) =\frac{1}{x (\overline{n}\cdot P_1)^2} \langle N_1 | 
\mathcal{B}_n^{\perp\mu a} \Bigl[\delta \Bigl( x_1 -
\frac{\overline{n}\cdot \mathcal{P}}{\overline{n}\cdot P_1}\Bigr)
\mathcal{B}_{n\mu}^{\perp a}\Bigr] |N_1\rangle.
\end{equation}
For $f_{g/N_2}$, $n$ and $\overline{n}$ are switched in Eq.~(\ref{defcg}). 

In terms of the collinear gluon distribution functions, the cross section is written as
\begin{eqnarray} \label{higgscon}
\sigma (pp\rightarrow HX) &=& \sigma_0\int dx_1 dx_2 H_{\mathrm{Higgs}}
(Q,\mu) S_{\mathrm{Higgs}} (z,\mu) f_{g/N_1} (x_1,\mu) f_{g/N_2} (x_2, \mu),
\end{eqnarray}
where the soft function $S_{\mathrm{Higgs}} (z,\mu)$ is defined as
\begin{equation}
S_{\mathrm{Higgs}} (z,\mu) = 
\frac{1}{N_c^2 -1} \langle 0| \mathrm{tr}\, \cy_n^{\dagger}
\cy_{\bar{n}} \delta \Bigl( 1-z  +\frac{2v\cdot \mathcal{R}}{Q}\Bigr) \cy_{\bar{n}}^{\dagger}
\cy_n |0\rangle.
\end{equation}
The Born cross section $\sigma_0$ is given by
\begin{equation}
\sigma_0 = \frac{2\pi  |C_t (m_t,\mu)|^2}{v^2 (N_c^2-1)},
\end{equation}
and $H_{\mathrm{Higgs}} (Q,\mu) = |C_H (Q,\mu)|^2$.

Here we invoke the relation Eq.~(\ref{injet}) to write the cross section as
\begin{eqnarray} \label{fachiggs}
\sigma (pp \rightarrow HX) /\sigma_0&=&   H_{\mathrm{Higgs}} (Q,\mu)
\int_{\tau}^1 \frac{dw}{w} \int_{\tau/w}^1 dx\phi_{g/N_1} (x,\mu)
\phi_{g/N_2} \Bigl(\frac{\tau}{xw},\mu\Bigr) \nonumber \\
&\times&\int_w^1 \frac{dz}{z} S_{\mathrm{Higgs}}(z,\mu) \int_{w/z}^1 
dt K_{gg} (t,\mu) K_{gg} \Bigl(\frac{w}{zt},\mu\Bigr) \nonumber \\
&=&H_{\mathrm{Higgs}} (Q,\mu) \int_{\tau}^1 \frac{dw}{w} W_{\mathrm{Higgs}}(w,\mu)
\int_{\tau/w}^1 dx \phi_{g/N_1} (x,\mu)
\phi_{g/N_2} \Bigl(\frac{\tau}{xw},\mu\Bigr) \nonumber \\
&=&H_{\mathrm{Higgs}} (Q,\mu) \int_{\tau}^1 \frac{dw}{w}W_{\mathrm{Higgs}} (w,\mu) 
\Phi_{gg} \Bigl(\frac{\tau}{w},\mu\Bigr),
\end{eqnarray}
where $W_{\mathrm{Higgs}}$ and $\Phi_{gg}$ are given as
\begin{eqnarray}  \label{defwhig}
W_{\mathrm{Higgs}} (w,\mu) &=& \int_w^1
\frac{dz}{z}  S_{\mathrm{Higgs}} (z,\mu) \int_{w/z}^1 dt K_{gg} (t,\mu)
K_{gg} \Bigl(\frac{w}{zt},\mu\Bigr), \nonumber \\ 
\Phi_{gg}
\Bigl(\frac{\tau}{w},\mu \Bigr) &=& \int_{\tau/w}^1dx\phi_{g/N_1}
(x,\mu) \phi_{g/N_2} \Bigl(\frac{\tau}{xw},\mu\Bigr). 
\end{eqnarray}
Eq.~(\ref{fachiggs}) is our new factorization theorem for the Higgs production near the 
partonic threshold.
 
\section{Soft function\label{softf}} 
In the traditional factorization scheme, the most problematic part is the soft function 
because it includes the IR and mixed divergences. Unless these divergences are 
removed, the soft function is not physical and the evolution via the renormalization group
equation is meaningless. Especially the real gluon emission in DY process has only IR 
divergences. From this section, the ingredients in the factorization 
formulae Eqs.~(\ref{fdyour}), and (\ref{f1our}) are computed at one loop. The UV 
and IR divergences appear unequivocally in the soft functions, and we compute the 
one-loop corrections to the soft functions.

\begin{figure}[b] 
\begin{center}
\includegraphics[height=3.5cm]{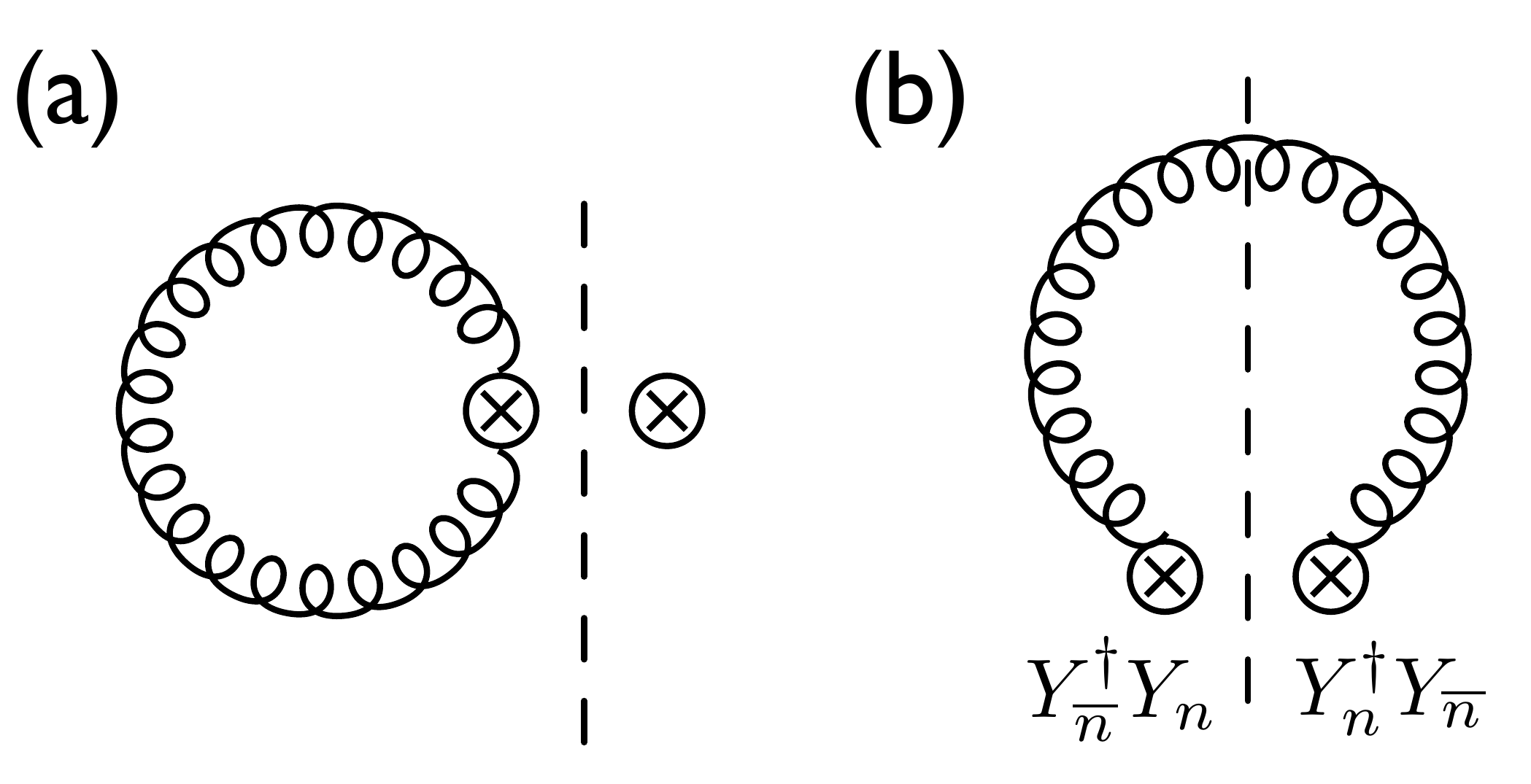}
\end{center}  
\vspace{-0.3cm}
\caption{Feynman diagrams for soft functions at one loop  
(a) virtual corrections and (b) real gluon emission. \label{softdisf}}
\end{figure}

In DY process, the soft function is given by
\begin{equation}  \label{softdy}
S_{\mathrm{DY}}  (z,\mu) = \frac{1}{N_c}
\langle 0|\mathrm{tr} Y_n^{\dagger} Y_{\bar{n}} \delta \Bigl( 1-z+\frac{2
v\cdot \mathcal{R}}{Q}\Bigr) Y_{\bar{n}}^{\dagger} Y_n|0\rangle, 
\end{equation}
The Feynman diagrams for the radiative
corrections of the soft function at one loop are given in Fig.~\ref{softdisf}.
Fig.~\ref{softdisf} (a) with its mirror image corresponds to the virtual correction, and Fig.~\ref{softdisf}(b) is the real gluon emission. The computation is performed by
employing the dimensional
regularization to regulate both the UV and IR divergences with the 
spacetime dimension $D=4-2\epsilon$ and the $\overline{\mathrm{MS}}$ scheme. 
The results with another
regularization scheme, in which the dimensional regularization is employed for the 
UV divergence, and the IR divergence is regulated by the offshellness, are presented
in Appendix A. 

The corresponding matrix elements are given as
\begin{eqnarray} \label{matdy}
M_{s,\mathrm{DY}}^a &=& -2ig^2 C_F \dimf \int \dfl \frac{1}{l^2 n\cdot l
\overline{n}\cdot l}  =-\frac{\alpha_s C_F}{2\pi} 
\Bigl( \frac{1}{\euv} -\frac{1}{\eir}\Bigr)^2 \delta (1-z), \nonumber \\
M_{s,\mathrm{DY}}^b&=& 4\pi g^2 C_F \dimf \int \dfl 
\frac{1}{n\cdot l \overline{n} \cdot l} \delta (l^2) \delta 
\Bigl(1-z-\frac{2v \cdot l}{Q} \Bigr)\nonumber \\
&=&\frac{\alpha_s C_F}{2\pi} \Bigl[ \delta (1-z) \Bigl(
\frac{1}{\eir^2} +\frac{2}{\eir} \ln \frac{\mu}{Q} +2
\ln^2 \frac{\mu}{Q} -\frac{\pi^2}{4}\Bigr)  \nonumber \\
&&-\frac{2}{(1-z)_+} \Bigl( \frac{1}{\eir} + 2\ln \frac{\mu}{Q}\Bigr)
+4\Bigl(\frac{\ln (1-z)}{1-z}\Bigr)_+\Bigr].
\end{eqnarray}
For the real gluon emission as in $M_{s,\mathrm{DY}}^b$, and in collinear contributions, 
$l_0>0$ is implied.
In this computation, the function $1/(1-z)^{1+\epsilon}$ appears. It diverges at $z=1$, and
the divergence is definitely of the IR origin. In terms of the plus distribution functions, it
can be expanded in powers of $\epsilon$ as
\begin{equation}
\frac{1}{(1-z)^{1+\epsilon}} =-\frac{1}{\epsilon} \delta (1-z) +\frac{1}{(1-z)_+} -
\epsilon \Bigl(\frac{\ln (1-z)}{1-z}\Bigr)_+ +\cdots.
\end{equation}
Here $M_{s,\mathrm{DY}}^b$ is the result with $Y_n$ ($Y_{\bar{n}}$) 
in the left- (right-) hand side in Fig.~\ref{softdisf} (b), and the result with $Y_{\bar{n}}^{\dagger}$
and $Y_n^{\dagger}$ should be included.  They are hermitian conjugates to each other.
The virtual correction has the UV and IR divergences, and 
the mixed divergence.  Note that there are only IR divergences in the real gluon 
emission. 

The total soft contribution for DY process at one loop is given as
\begin{eqnarray} \label{sdyone}
S_{\mathrm{DY}}^{(1)} (z) &=& 2  
(M_{s,\mathrm{DY}} ^a +M_{s,\mathrm{DY}} ^b)  \nonumber \\
&=&\frac{\alpha_s C_F}{\pi} \Bigl[ \delta (1-z) \Bigl(
-\frac{1}{\euv^2}+\frac{2}{\euv\eir} +\frac{1}{\eir} \ln \frac{\mu^2}{Q^2} 
+\frac{1}{2}\ln^2 \frac{\mu^2}{Q^2} -\frac{\pi^2}{4}\Bigr)  \nonumber \\
&&-\frac{2}{(1-z)_+} \Bigl( \frac{1}{\eir} + \ln \frac{\mu^2}{Q^2}\Bigr)
+4\Bigl(\frac{\ln (1-z)}{1-z}\Bigr)_+\Bigr].
\end{eqnarray}
In Eq.~(\ref{sdyone}), the finite terms are the same as the result in Ref.~\cite{Korchemsky:1993uz}. 
But the divergent terms are mixture of IR  and UV divergences
contrary to the argument in Ref.~\cite{Korchemsky:1993uz}, in which they claim that the 
IR divergences are cancelled due to the KLN theorem. Our explicit calculation shows that it 
is not true. The KLN theorem holds when there is no kinematic constraint on the real gluon 
emission. However, as can be seen in $M_{s,\mathrm{DY}}^b$ for the real gluon emission, 
there is a constraint for the soft momentum of the real gluon specified by the delta function, 
which is different from the virtual corrections. Therefore the cancellation of the IR divergence
is bound to be incomplete.

The soft function for the Higgs production $S_{\mathrm{Higgs}}(z)$ is defined as
\begin{equation} \label{softhiggs}
S_{\mathrm{Higgs}} (z) = \frac{1}{N_c^2 -1} \langle 0| \mathrm{tr} \cy_n^{\dagger} 
\cy_{\bar{n}} \delta \Bigl( 1-z+\frac{2v\cdot \mathcal{R}}{Q}\Bigr) \cy_{\bar{n}}^{\dagger} 
\cy_n |0\rangle,
\end{equation}
where $Q=m_H$, and the soft Wilson lines $\cy_n$, and $\cy_{\bar{n}}$ are in the adjoint
representation. $S_{\mathrm{Higgs}} (z)$ is normalized to $\delta (1-z)$ at tree level. 
In this expression, Eq.~(\ref{softhiggs}) resembles Eq.~(\ref{softdy}), and the only difference
is the representation of the soft Wilson lines. This means that the radiative corrections for
$S_{\mathrm{Higgs}}$ is the same as $S_{\mathrm{DY}}$ except the color factors. By explicit 
computation, it is true and the corresponding radiative corrections for $S_{\mathrm{Higgs}}$
are obtained from Eq.~(\ref{matdy}) by replacing $C_F$ by $C_A$. Accordingly, the
one-loop correction for $S_{\mathrm{Higgs}}$ is given by
\begin{eqnarray} \label{shone}
S_{\mathrm{Higgs}}^{(1)} (z) &=&  \frac{\alpha_s C_A}{\pi} \Bigl[ \delta (1-z) \Bigl(
-\frac{1}{\euv^2}+\frac{2}{\euv\eir} +\frac{1}{\eir} \ln \frac{\mu^2}{Q^2} 
+\frac{1}{2}\ln^2 \frac{\mu^2}{Q^2} -\frac{\pi^2}{4}\Bigr)  \nonumber \\
&&-\frac{2}{(1-z)_+} \Bigl( \frac{1}{\eir} + \ln \frac{\mu^2}{Q^2}\Bigr)
+4\Bigl(\frac{\ln (1-z)}{1-z}\Bigr)_+\Bigr].
\end{eqnarray}

The soft function $S_{\mathrm{DIS}}$ in DIS near the endpoint is given by 
\begin{equation} \label{softdis}
S_{\mathrm{DIS}} (z) =\frac{1}{N_c} \langle 0|\mathrm{tr}
\Bigl[ Y_n^{\dagger} \tilde{Y}_{\bar{n}} \delta \Bigl( 1-z
+\frac{\overline{n}\cdot \mathcal{R}}{Q}\Bigr) \tilde{Y}_{\bar{n}}^{\dagger} Y_n
\Bigr] |0\rangle. 
\end{equation}
The Feynman diagrams for the one-loop correction is shown in Fig.~\ref{softdisf}, 
except that $Y_{\bar{n}}$ is replaced by $\tilde{Y}_{\bar{n}}$.
The matrix elements are given as
\begin{eqnarray} 
M_{s,\mathrm{DIS}}^a &=& -\frac{\alpha_s C_F}{2\pi} \delta (1-z)
\Bigl(\frac{1}{\euv} -\frac{1}{\eir}\Bigr)^2, \nonumber \\ 
M_{s,\mathrm{DIS}}^b &=&
-\frac{\alpha_s C_F}{2\pi} \Bigl( \frac{1}{\euv} -\frac{1}{\eir}\Bigr)
\Bigl[\Bigl(\frac{1}{\eir} + \frac{1}{2}\ln \frac{\mu^2}{Q^2} \Bigr) \delta (1-z)
-\frac{1}{(1-z)_+} \Bigr]. 
\end{eqnarray}
Unlike DY process, the real gluon emission contains UV divergence. 
The total soft contribution in DIS at one loop is given by 
\begin{eqnarray} \label{sdisone}
S_{\mathrm{DIS}}^{(1)} (z) &=& 2  (M_{s,\mathrm{DIS}}^a 
+M_{s,\mathrm{DIS}}^b) \\ &=& \frac{\alpha_s
C_F}{\pi} \Bigl( \frac{1}{\euv} -\frac{1}{\eir}\Bigr)
\Bigl[\Bigl(-\frac{1}{\euv} -\frac{1}{2}\ln \frac{\mu^2}{Q^2}\Bigr) \delta (1-z)
+\frac{1}{(1-z)_+} \Bigr]. \nonumber 
\end{eqnarray}

All the soft functions contain IR, UV divergences as well as mixed 
divergence. Therefore they are not physically meaningful as they are. As claimed, 
only after the initial-state jet function is added, the IR and mixed
divergences disappear.  

\section{Collinear distribution functions and PDF \label{colli}} 
\subsection{Collinear quark distribution function and PDF}
\begin{figure}[b] 
\begin{center}
\includegraphics[height=3.5cm]{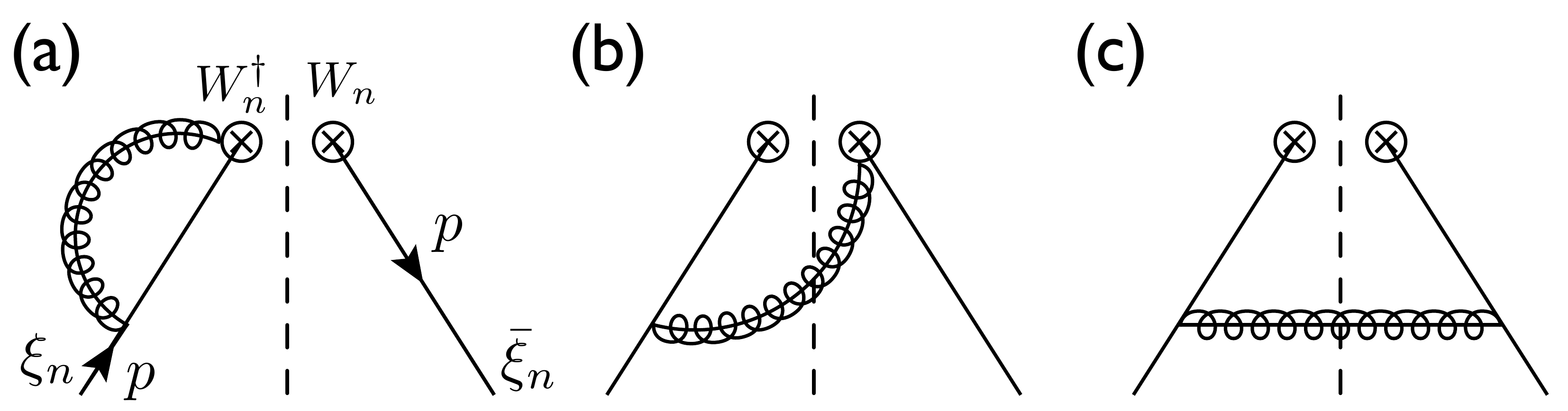}
\end{center}  
\vspace{-0.3cm}
\caption{Feynman diagrams for collinear functions and PDF at one loop  
(a) virtual corrections, (b) and (c) real gluon emission. The mirror images of (a)
and (b) are omitted.\label{cof}}
\end{figure}

The radiative corrections of the collinear quark distribution function, defined as 
\begin{equation} f_{q/N} (x,\mu) = \langle N|\overline{\chi}_n
\frac{\FMslash{\overline{n}}}{2} \delta (\overline{n}\cdot P x -\overline{n}\cdot \mathcal{P})
\chi_n |N\rangle,
\end{equation} 
will be computed explicitly at one loop.
The Feynman diagrams for the collinear
function at one loop are shown in Fig.~\ref{cof}. Fig.~\ref{cof} (a) is the
virtual correction, and Fig.~\ref{cof} (b) and (c) are real gluon emissions. 
The mirror images of Fig.~\ref{cof} (a) and (b) are omitted, which are
given by the hermitian conjugates of the original diagrams. 

The matrix elements are written as 
\begin{eqnarray} \label{colmat}
M_a &=& 2ig^2 C_F \dimf \delta (1-x) \int \dfl 
\frac{\overline{n}\cdot (p-l)}{l^2 (l-p)^2 (\overline{n}\cdot l +\delta_1)}, \nonumber \\ 
M_b &=& -4\pi g^2 C_F \dimf
\int \dfl \frac{\overline{n}\cdot (p-l)}{(l-p)^2 (\overline{n}\cdot l
+\delta_1)} \delta \Bigl(1-x-\frac{\overline{n}\cdot l}{\overline{n}\cdot p}\Bigr)
\delta (l^2), \nonumber \\
M_c &=& 2\pi g^2 C_F \dimf
(D-2) \int \dfl \frac{\mathbf{l}_{\perp}^2}{[(l-p)^2]^2} 
\delta \Bigl(1-x-\frac{\overline{n}\cdot l}{\overline{n}\cdot p}\Bigr)\delta (l^2).
\end{eqnarray}
These are computed using the dimensional regularization for both UV and IR divergences
with $p^2 =0$ in the $\overline{\mathrm{MS}}$ scheme. The results 
with the dimensional regularization for the UV divergence,
and the nonzero $p^2$ for the IR divergence are presented in Appendix B. 

The results of the computation are given by 
\begin{eqnarray} \label{naiveco}
M_a &=& \frac{\alpha_s C_F}{2\pi} \delta (1-x) \Bigl( \frac{1}{\euv}
-\frac{1}{\eir}\Bigr) \Bigl(1+\ln \frac{\delta_1}{\overline{n}\cdot p} \Bigr),
\nonumber \\ 
M_b &=& \frac{\alpha_s C_F}{2\pi}   \Bigl(
\frac{1}{\euv} -\frac{1}{\eir}\Bigr) \Bigl( -\delta (1-x) \ln
\frac{\delta_1}{\overline{n}\cdot p}+ \frac{x}{(1-x)_+}\Bigr), \nonumber \\ M_c
&=&\frac{\alpha_s C_F}{2\pi} (1-x) \Bigl( \frac{1}{\euv} -\frac{1}{\eir}\Bigr).
\end{eqnarray}
The rapidity regulator $\delta_1$ is employed in the computation.  As can be seen,
$M_a$ and $M_b$ depend on this regulator $\delta_1$, but this dependence is 
cancelled in the sum $M_a+M_b$ even without the zero-bin subtraction. This is to be
contrasted with the transverse-momentum-dependent collinear function, where 
the cancellation of $\delta_1$ is achieved only after the zero-bin subtraction
\cite{Chay:2012mh}.

In computing the zero-bin contributions, we neglect all the components
of the loop momentum $l$ compared to $\overline{n}\cdot p$.  
From Eq.~(\ref{colmat}), they are  given as
\begin{eqnarray}
M_a^{(0)} &=& -2ig^2 C_F \dimf
\delta (1-x) \int \dfl \frac{1}{l^2 n\cdot l (\overline{n}\cdot l
+\delta_1)} \nonumber \\
&=& -\frac{\alpha_s C_F}{2\pi} \delta (1-x) \Bigl( \frac{1}{\euv}
-\frac{1}{\eir}\Bigr) \Bigl( \frac{1}{\euv} - \ln \frac{\delta_1}{\mu}\Bigr),
\nonumber \\ 
M_{b,\mathrm{s}}^{(0)} &=& 4\pi g^2 C_F \dimf
\int \dfl\frac{1}{n\cdot l(\overline{n}\cdot l
+\delta_1)} \delta \Bigl(1-x-\frac{\overline{n}\cdot l}{\overline{n}\cdot p}\Bigr)
\delta (l^2) \nonumber \\
&=&\frac{\alpha_s C_F}{2\pi} \Bigl( \frac{1}{\euv}
-\frac{1}{\eir}\Bigr) \Bigl( -\delta (1-x) \ln \frac{\delta_1}{\overline{n}\cdot
p} +\frac{1}{(1-x)_+}\Bigr), \nonumber \\ 
M_c^{(0)} &=&0.
\end{eqnarray}
The zero-bin contribution $M_c^{(0)}$ is subleading and is neglected.
Note that the distinction between the soft and usoft  zero-bin contributions appears in
$M_{b,\mathrm{s}}^{(0)}$. In \one, the soft momentum
is of order $Q (1-x)$, which is of the same order as the loop momentum 
$\overline{n} \cdot l$ in the zero-bin contribution. 
However, there is no distinction between the soft and usoft contributions in 
$M_a^{(0)}$ because the integral remains the same irrespective of the size of the loop
momentum. $M_c^{(0)}$ can also be neglected in the usoft limit. Therefore the 
soft and usoft zero-bin contributions are different only in $M_b$, which corresponds
to the real gluon emission with soft or usoft momentum.  The nontrivial existence of the zero-bin
subtraction in \one\ and the absence in \two\ were also pointed out  when
the quark beam function was considered \cite{Stewart:2010qs}.

The collinear part with the soft zero-bin subtraction is written as
\begin{eqnarray} \label{coldist}
\tilde{M}_a &=& M_a -M_a^{(0)} = \frac{\alpha_s C_F}{2\pi}
\delta (1-x) \Bigl( \frac{1}{\euv} -\frac{1}{\eir}\Bigr) \Bigl( \frac{1}{\euv}
+1+\ln \frac{\mu}{\overline{n}\cdot p}\Bigr), \nonumber \\ 
\tilde{M}_b &=& M_b
-M_{b,\mathrm{s}}^{(0)} =-\frac{\alpha_s C_F}{2\pi} \Bigl( \frac{1}{\euv}
-\frac{1}{\eir}\Bigr), \nonumber \\ 
\tilde{M}_c &=& M_c = \frac{\alpha_s
C_F}{2\pi} (1-x) \Bigl( \frac{1}{\euv} -\frac{1}{\eir}\Bigr). 
\end{eqnarray}
Note that various combinations are independent of the rapidity regulator $\delta_1$
though individual diagrams depend on it. As was first noted, the sum of naive collinear
contributions is independent of $\delta_1$.  So is the sum of soft zero-bin contributions,
hence the true collinear contribution with the zero-bin subtraction. It is also true that
the usoft zero-bin contribution is also independent of $\delta_1$, in fact, it vanishes.
It turns out that the same result is obtained without
introducing $\delta_1$ in the beginning. But $\delta_1$ is included in the calculation
to show how the  cancellation occurs explicitly, and this method has been used 
also in calculating the transverse-momentum-dependent collinear distribution function 
\cite{Chay:2012mh}.  

In order to obtain the collinear distribution functions from the collinear part in 
Eq.~(\ref{coldist}) near the endpoint, we take the limit $x\rightarrow 1$ except in the singular
functions. The remaining terms are in powers of $1-x$, and they are neglected near the endpoint.
We can take the limit $x\rightarrow 1$ from the beginning of the computation, 
and the result is the same.
Here we take the endpoint limit later so that the comparison with the QCD result is transparent, and
this calculation can be performed both in the hadronic and the partonic thresholds.

The collinear distribution function at one loop near the endpoint is given by 
\begin{eqnarray} 
f^{(1)}_{q/N} (x,\mu)
&=& 2  (\tilde{M}_a +\tilde{M}_b) + \tilde{M}_c -\frac{\alpha_s C_F}{4\pi} \Bigl(
\frac{1}{\euv} -\frac{1}{\eir}\Bigr)  \delta (1-x)\nonumber \\ 
&=&\frac{\alpha_s C_F}{2\pi}
\Bigl( \frac{1}{\euv} -\frac{1}{\eir}\Bigr) \Bigl[\Bigl( \frac{2}{\euv}
+\frac{3}{2} +2\ln \frac{\mu}{\overline{n}\cdot p} \Bigr) \delta (1-x) -2\Bigr], 
\end{eqnarray}
where the last term in the first line comes from the
self-energy of the fermion field $\xi$. The radiative correction $f^{(1)}_{q/N}$
contains the UV and IR divergences, furthermore there is the mixing of 
the UV and IR divergences. Due to the IR and mixed divergences, it is not 
appropriate to discuss the evolution of the collinear distribution function. Our factorization 
procedure removes the mixing of the UV and IR divergences as will be shown later.

In $\mathrm{SCET}_{\mathrm{II}}$, the zero-bin subtraction must be usoft, and
the only difference from the soft zero-bin subtraction appears in $\tilde{M}_b$.
The usoft zero-bin contribution for $M_{b,\mathrm{us}}^{(0)}$ is given by
\begin{eqnarray} 
M_{b,\mathrm{us}}^{(0)} &=&
4\pi g^2 C_F \dimf \delta (1-x) \int \frac{d^D l}{(2\pi)^D} 
\frac{\delta (l^2) }{n\cdot l(\overline{n}\cdot l+\delta_1)}   \nonumber \\
&=& \frac{\alpha_s C_F}{2\pi} \delta
(1-x) \Bigl( \frac{1}{\euv} -\frac{1}{\eir}\Bigr) \Bigl( \frac{1}{\euv} -\ln
\frac{\delta_1}{\mu}\Bigr), 
\end{eqnarray}
and the corresponding collinear contribution is given by 
\begin{equation}
\tilde{M}_{b,\mathrm{us}} = M_b - M_{b,\mathrm{us}}^{(0)} = \frac{\alpha_s
C_F}{2\pi} \Bigl( \frac{1}{\euv} -\frac{1}{\eir}\Bigr) \Bigl[\delta (1-x)
\Bigl(-\frac{1}{\euv} -\ln \frac{\mu}{\overline{n}\cdot p}\Bigr)
+\frac{x}{(1-x)_+}\Bigr]. 
\end{equation} 
Note that $M_{a,\mathrm{us}} = M_{a,\mathrm{s}}= -M_{b,\mathrm{us}}$, hence
$M_{a,\mathrm{us}}+M_{b,\mathrm{us}}=0$, meaning that there is no
usoft zero-bin contributions in the PDF.  

The PDF $\phi^{(1)}_{q/N} (x,\mu)$ at one loop near the endpoint is given by 
\begin{eqnarray} \label{phione}
\phi^{(1)}_{q/N} (x,\mu)  &=& 2   (\tilde{M}_a +\tilde{M}_{b,\mathrm{us}} ) 
+ \tilde{M}_c -\frac{\alpha_s C_F}{4\pi} \Bigl(
\frac{1}{\euv} -\frac{1}{\eir}\Bigr) \delta (1-x) \nonumber \\ 
&=&\frac{\alpha_s C_F}{2\pi} \Bigl( \frac{1}{\euv} -\frac{1}{\eir}\Bigr) 
\Bigl(\frac{3}{2} \delta
(1-x) +\frac{2}{(1-x)_+}\Bigr). 
\end{eqnarray}
Unlike $f^{(1)}_{q/N}$, the radiative correction $\phi^{(1)}_{q/N}$ does not 
involve the
mixed divergence. The result is the same as in full QCD. The IR divergence is
absorbed in the nonperturbative part of $\phi_{q/N}$, and the UV divergent part
yields the anomalous dimension of the PDF, which governs the renormalization
group behavior.

The fact that the usoft zero-bin contributions in the PDF vanish is responsible for
why full QCD results can be obtained from naive collinear contribution only. 
Of course, the full QCD results hold away from the endpoint region, and near the
endpoint.  It can be clearly explained in SCET. Away from the endpoint region, 
there is no intermediate scale and there is no need to construct effective theories
successively. And there are no delta functions in the soft functions in 
Eqs.~(\ref{softdy}) and (\ref{softdis}). Then the soft functions are just one, meaning that
there are no soft contributions to all orders in $\alpha_s$. And 
the radiative corrections of the collinear part at low energy scale are exactly those
of the PDF given by Eq.~(\ref{phione}). However, there is no way to avoid
the steps described above near the endpoint region. 

The relation between the collinear function $f_{q/N}$ and the PDF $\phi_{q/N}$ 
is given by
\begin{equation} 
f_{q/N}(x,\mu) =\int_x^1 \frac{dz}{z} K_{qq}(z,\mu) \phi_{q/N}
\Bigl(\frac{x}{z},\mu\Bigr), 
\end{equation}
where $K_{qq}(z,\mu)$ is the difference between the zero-bin subtractions
in the collinear matrix element between $\mathrm{SCET}_{\mathrm{I}}$
and $\mathrm{SCET}_{\mathrm{II}}$. At one loop, it  is given by 
\begin{eqnarray}  \label{kqq}
K_{qq}^{(1)}(z,\mu) &=&
2   (-M_{b,\mathrm{s}}^{(0)} +M_{b,\mathrm{us}}^{(0)}) \\ 
&=&\frac{\alpha_s C_F}{\pi} \Bigl(
\frac{1}{\euv} -\frac{1}{\eir}\Bigr) \Bigl[\delta (1-z) \Bigl(\frac{1}{\euv}
+\ln \frac{\mu}{\overline{n}\cdot p}\Bigr) -\frac{1}{(1-x)_+}\Bigr]. \nonumber
\end{eqnarray}
We stress again, $K_{qq}$ is not the matching coefficient between \one\ and \two.
It contains IR divergence, and the mixed divergence too. There also appear IR
and mixed divergences in the soft function. 
Combining these two, there will be neither
IR nor mixed divergence, as is explicitly shown at one loop.

\subsection{Collinear gluon distribution function and PDF}

\begin{figure}[t] 
\begin{center}
\includegraphics[height=3.5cm]{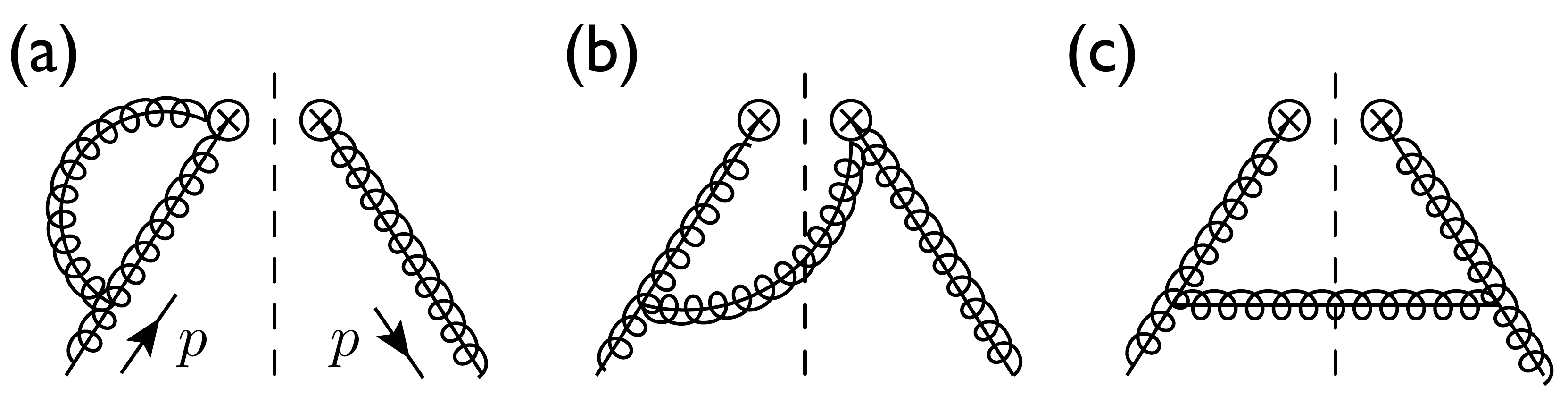}
\end{center}  
\vspace{-0.3cm}
\caption{Feynman diagrams for collinear gluon distribution functions and PDF at one loop  
(a) virtual corrections, (b) and (c) real gluon emission. The mirror images of (a) and 
(b) are omitted.\label{cofg}}
\end{figure}

The collinear gluon distribution function is defined as
\begin{equation}
f_{g/N} (x,\mu) = \frac{1}{x \overline{n}\cdot P} \langle N(P)| \mathcal{B}_n^{\perp\mu a}
\delta (x \overline{n}\cdot P - \overline{n}\cdot \mathcal{P}) \mathcal{B}_{n\mu}^{\perp a}
|N(p)\rangle,
\end{equation}
which is normalized to $f_{g/N} (x) = \delta (1-x)$ at tree level.  
The Feynman diagrams for the collinear gluon distribution function at one loop is shown in 
Fig.~\ref{cofg}. 
%The field $\mathcal{B}_n^{\perp\mu a}$ can be expanded as
%\begin{eqnarray}
%\mathcal{B}_n^{\perp\mu a} &=& \overline{n}\cdot \mathcal{P} A_{\perp a}^{\mu} 
%-\mathcal{P}_{\perp} ^{\mu} \overline{n}\cdot A_{na} \nonumber \\
%&+& igf^{abc} \Bigl( \Bigl[\mathcal{P}_{\perp}^{\mu} \overline{n}\cdot A_b -\overline{n}
%\cdot \mathcal{P}A_{\perp b}^{\mu} \Bigr] \frac{1}{\overline{n}\cdot \mathcal{P}
%+\delta_1} \overline{n} \cdot A_{n c} + \overline{n}\cdot A_{nb} A_{\perp c}^{\mu} \Bigr)
%\nonumber \\
%&+& g^2 f^{bcd} f^{abe} \overline{n}\cdot A_{nc} A_{\perp d}^{\mu} 
%\frac{1}{\overline{n}\cdot \mathcal{P} +\delta_1} \overline{n}\cdot A_{ne} +\cdots,
%\end{eqnarray}
%from which the Feynman rules for the vertices in Fig.~\ref{cofg} can be derived.

The matrix elements in the background gauge are given as
\begin{eqnarray}
M_a &=& 2ig^2 C_A\overline{n}\cdot p \delta (1-x) \dimf \int \dfl  \frac{1}{l^2 (l+p)^2}
\Bigl[\frac{1}{\overline{n}\cdot l -\delta_1} -\frac{1}{\overline{n}\cdot (l+p) +\delta_1} \Bigr]
\nonumber \\
&=& \frac{\alpha_s C_A}{2\pi} \delta (1-x) \Bigl( \frac{1}{\euv} -\frac{1}{\eir}\Bigr) 
\ln \frac{\delta_1}{\overline{n}\cdot p}, \nonumber \\
M_b &=&  \frac{2\pi g^2 C_A}{x\overline{n}\cdot p} \dimf \int \dfl \delta (l^2) \delta 
\Bigl( \overline{n}\cdot l -(1-x) \overline{n}\cdot p\Bigr) \nonumber \\
&&\times \frac{[\overline{n} \cdot (l-p)]^2}{(l-p)^2} \Bigl[
\frac{\overline{n}\cdot (p-2l)}{\overline{n}\cdot p} -
\frac{2\overline{n}\cdot p}{\overline{n}\cdot l +\delta_1} \Bigr] \nonumber \\
&=& \frac{\alpha_s C_A}{2\pi} \Bigl( \frac{1}{\euv} -\frac{1}{\eir}\Bigr) \Bigl(
-\delta(1-x) \ln \frac{\delta_1}{\overline{n}\cdot p} +\frac{x}{(1-x)_+}+ x(1-x) -\frac{x}{2}
\Bigr), \nonumber \\ 
M_c &=& \frac{8\pi g^2 C_A}{x\overline{n}\cdot p} \dimf \int \dfl \delta (l^2) \delta 
\Bigl( \overline{n}\cdot l -(1-x) \overline{n}\cdot p\Bigr)  \frac{1}{(n\cdot l)^2} 
\Bigl( \frac{x^2}{2} n\cdot l \overline{n}\cdot p -\mathbf{l}_{\perp}^2 \Bigr) \nonumber \\
&=& \frac{\alpha_s C_A}{\pi} \Bigl( \frac{1}{\euv} -\frac{1}{\eir}\Bigr)
\Bigl( \frac{x}{2} +\frac{1-x}{x}\Bigr).
\end{eqnarray}

The soft zero-bin contributions are given as
\begin{eqnarray}
M_a^{(0)} &=& -\frac{\alpha_s C_A}{2\pi} \Bigl( \frac{1}{\euv} -\frac{1}{\eir}\Bigr) 
\Bigl( \frac{1}{\euv} -\ln \frac{\delta_1}{\mu}\Bigr) \delta (1-x), \nonumber \\
M_{b,\mathrm{s}}^{(0)} &=& \frac{\alpha_s C_A}{2\pi}  
\Bigl( \frac{1}{\euv} -\frac{1}{\eir}\Bigr) 
\Bigl( -\delta (1-x) \ln \frac{\delta_1}{\overline{n}\cdot p} + \frac{1}{(1-x)_+} \Bigr),
\nonumber \\
M_c^{(0)} &=&0. 
\end{eqnarray}
The usoft zero-bin contribution only differs in the real gluon emission of $M_b$, and it is 
given as
\begin{equation}
M_{b,\mathrm{us}}^{(0)} = \frac{\alpha_s C_A}{2\pi} \delta (1-x) 
\Bigl( \frac{1}{\euv} -\frac{1}{\eir}\Bigr)  \Bigl(\frac{1}{\euv} -\ln \frac{\delta_1}{\mu}\Bigr).
\end{equation}
Therefore the matrix elements of the collinear gluon distribution function at one loop with 
the soft zero-bin subtraction are given by
\begin{eqnarray}
\tilde{M}_a &=& M_a -M_a^{(0)} = \frac{\alpha_s C_A}{2\pi} 
\Bigl( \frac{1}{\euv} -\frac{1}{\eir}\Bigr)  \Bigl( \frac{1}{\euv}
+\ln \frac{\mu}{\overline{n}\cdot p}\Bigr) \delta (1-x), \nonumber \\
\tilde{M}_{b,\mathrm{s}} &=& M_b -M_{b,\mathrm{s}} {(0)} = 
\frac{\alpha_s C_A}{2\pi} \Bigl( \frac{1}{\euv} -\frac{1}{\eir}\Bigr)
\Bigl(x(1-x) -\frac{x}{2} -1\Bigr), \nonumber \\
\tilde{M}_c &=& M_c,  
\end{eqnarray}
and the usoft zero-bin subtraction $\tilde{M}_{b,\mathrm{us}}$ is given by
\begin{eqnarray}
\tilde{M}_{b,\mathrm{us}}&=& M_b -M_{b,\mathrm{us}}^{(0)}  \\
&=& \frac{\alpha_s C_A}{2\pi} \Bigl( \frac{1}{\euv} -\frac{1}{\eir}\Bigr) \Bigl[ \Bigl(
-\frac{1}{\euv} -\ln \frac{\mu}{\overline{n}\cdot p}\Bigr) \delta (1-x) +\frac{x}{(1-x)_+}
+x(1-x) -\frac{x}{2}\Bigr]. \nonumber 
\end{eqnarray}

Combining all the ingredients, the collinear gluon distribution function at one 
loop near the endpoint taking the limit $x\rightarrow 1$ is written as
\begin{eqnarray} \label{cogd}
f_{g/N}^{(1)} (x,\mu)&=& 2    (\tilde{M}_a +\tilde{M}_{b,\mathrm{s}}) +M_c +
\frac{\alpha_s \beta_0}{4\pi} \delta (1-x) \Bigl( \frac{1}{\euv} -\frac{1}{\eir}\Bigr) \\
&=& \frac{\alpha_s C_A}{\pi} \Bigl( \frac{1}{\euv} -\frac{1}{\eir}\Bigr) \Bigl[ \Bigl(
\frac{1}{\euv} +\ln \frac{\mu}{\overline{n}\cdot p}  +\frac{11}{12} -\frac{n_f}{6N_c}
\Bigr) \delta (1-x) -1 \Bigr],  \nonumber
\end{eqnarray}
where the last term in the first line is the self-energy correction of the gluon field in the
background gauge, and $\beta_0$ is given by
\begin{equation}
\beta_0 = \frac{11}{3} N_c -\frac{2}{3}n_f.
\end{equation}
The gluon PDF $\phi_{g/N}^{(1)}$ near the endpoint is obtained  by replacing 
$\tilde{M}_{b,\mathrm{s}}$ with $\tilde{M}_{b,\mathrm{us}}$ in Eq.~(\ref{cogd}), and is given by
\begin{equation}
\phi_{g/N}^{(1)} (x,\mu) =\frac{\alpha_s C_A}{\pi} \Bigl( \frac{1}{\euv} -\frac{1}{\eir}\Bigr)
\Bigl[ \Bigl(\frac{11}{12} -\frac{n_f}{6N_c}\Bigr) \delta (1-x)  +\frac{1}{(1-x)_+}\Bigr].
\end{equation}
The one-loop correction to $\phi_{g/N}^{(1)}$ is the same as the result in full QCD for 
the same reason as in the case of the quark PDF. The initial-state jet function at one loop
is given by
\begin{eqnarray} \label{kgg}
K_{gg}^{(1)} (x,\mu) &=& 2   (-M_{b,\mathrm{s}}^0 +M_{b,\mathrm{us}}^0) \\
&=& \frac{\alpha_s C_A}{\pi} \Bigl( \frac{1}{\euv} -\frac{1}{\eir}\Bigr) \Bigl[  
\Bigl(\frac{1}{\euv} +\ln \frac{\mu}{\overline{n}
\cdot p} \Bigr) \delta (1-x) - \frac{1}{(1-x)_+}\Bigr]. \nonumber 
\end{eqnarray}
Compared to the initial-jet function with quarks in Eq.~(\ref{kqq}), $K_{gg}^{(1)}$ is the same
except the color factor, satisfying $K_{qq}^{(1)} (x) /C_F = K_{gg}^{(1)} (x)/C_A$.

\section{Final-state jet function\label{jetf}} 
The final-state jet function for DIS in the $\overline{n}$ direction is defined as 
\begin{equation} \sum_{X_{\bar{n}} }
\chi_{\bar{n}} |X_{\bar{n}} \rangle \langle X_{\bar{n}} | \overline{\chi}_{\bar{n}}
=\frac{\FMslash{\overline{n}}}{2} \int\frac{d^4 p_X}{(2\pi)^4} \overline{J}(p_X).
\end{equation}
The jet function $\overline{J} (n\cdot p_X)$ is a function of $n\cdot p_X$ only. 
As defined in Eq.~(\ref{js}), the radiative corrections to  
$J(z,\mu) =Q\overline{J}(Q(1-z)) /(2\pi)$ will be computed. It has been computed 
to two-loop order \cite{Becher:2006qw}, 
but we will explicitly present the calculation at one loop to show how it can be 
computed like the collinear distribution
function. However, the final states are not on the mass shell $p_X^2 \sim Q^2 (1-z)$, 
the IR divergence is regulated by the offshellness $p_X^2$. In the intermediate calculation,
the IR poles appear, but in the final sum they cancel.
 
\begin{figure}[b] 
\begin{center}
\includegraphics[height=3.5cm]{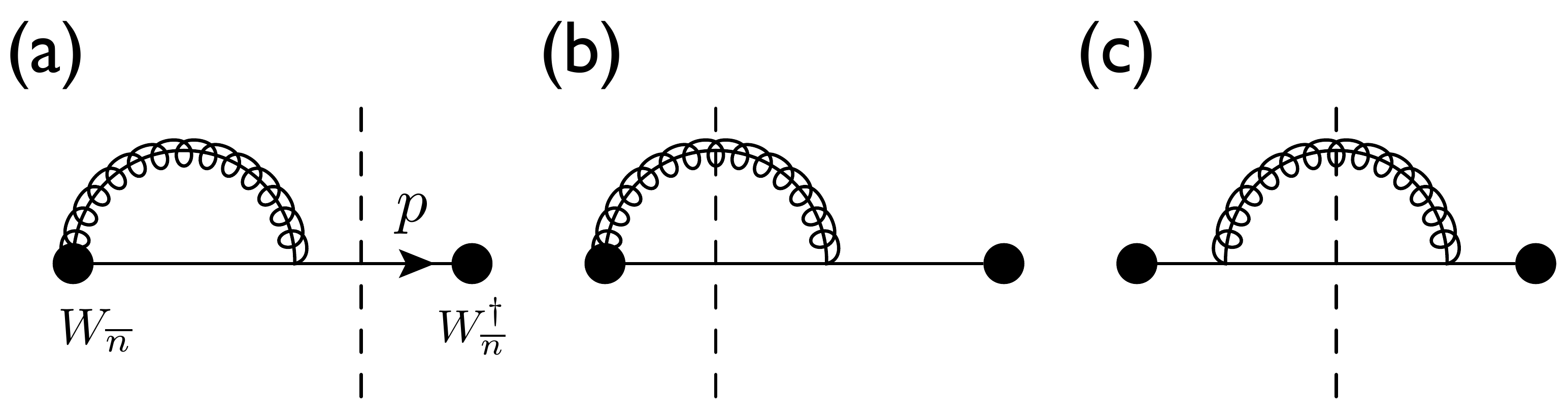}
\end{center}  
\vspace{-0.3cm}
\caption{Feynman diagrams for the final-state jet functions in DIS  at one loop  
(a) virtual corrections, (b) and (c) real gluon emission. The mirror images of (a)
and (b) are omitted.\label{jetdis}}
\end{figure}
The Feynman diagrams for the final-state jet function at one loop are shown
in Fig.~\ref{jetdis}. Fig.~\ref{jetdis} (a) and its mirror image are the virtual
corrections, and Fig.~\ref{jetdis} (b) and (c) are the real gluon emissions. 
The virtual correction of the fermion self energy is omitted in the figure, but
is added separately in the final calculation. Their matrix elements are given as 
\begin{eqnarray}  \label{jetco}
M_a &=& -2ig^2 C_F \delta (1-z)  \dimf \int \dfl
\frac{n\cdot (l+p)}{l^2 (l+p)^2(n\cdot l +\delta_2)} \nonumber \\
&=&\frac{\alpha_s C_F}{2\pi} \delta (1-z) \Bigl(
\frac{1}{\euv} -\frac{1}{\eir}\Bigr) \Bigl(1+\ln
\frac{-\delta_2}{n\cdot p} \Bigr),  \nonumber \\
M_b &=& 8\pi^2 g^2 C_F  \dimf \frac{Q^2}{p^2}
\int \dfl  \frac{n\cdot (p-l)}{(n\cdot l
-\delta_2)} \delta (l^2) \delta \Bigl((l-p)^2 \Bigr)\nonumber \\
&=& \frac{\alpha_s C_F}{2\pi}\Bigl\{ \delta (1-z) \Bigl[ \Bigl(\frac{1}{\eir}
+2\ln \frac{\mu}{n\cdot p}\Bigr) 
\Bigl( 1+\ln \frac{-\delta_2}{n\cdot p}\Bigr) +2
-\frac{\pi^2}{3} -\frac{1}{2}\ln^2 \frac{-\delta_2}{n\cdot p} \nonumber \\
&&-\frac{1}{(1-z)_+} \Bigl(1+\ln \frac{-\delta_2}{n\cdot p}\Bigr) \Bigr\}, 
\nonumber \\
M_c &=& 2\pi g^2 C_F (D-2) \frac{Q^3}{(p^2)^2}  \dimf
\int \dfl  \frac{\mathbf{l}_{\perp}^2}{n\cdot (p-l)}
\delta (l^2) \delta \Bigl((l-p)^2 \Bigr) \nonumber \\
&=& \frac{\alpha_s C_F}{4\pi} \Bigl[ -\delta (1-z) \Bigl( \frac{1}{\eir}+
2\ln \frac{\mu}{n\cdot p} +1\Bigr) +\frac{1}{(1-z)_+}\Bigr].
\end{eqnarray}

To be consistent with the idea of the zero-bin subtraction, the soft contributions
should be subtracted from the naive collinear calculation. The zero-bin 
contributions in this case are easy to deduce, and the soft zero-bin
subtraction is the appropriate procedure. The zero-bin contributions from 
Eq.~(\ref{jetco}) are given as
\begin{eqnarray}
M_a^{(0)} &=& -\frac{\alpha_s C_F}{2\pi} \delta (1-z)\Bigl(\frac{1}{\euv}
-\frac{1}{\eir}\Bigr) \Bigl( 1+\ln \frac{-\delta_2}{n\cdot p}\Bigr), \nonumber \\
M_b^{(0)} &=& \frac{\alpha_s C_F}{2\pi} \Bigl[ \delta (1-z)  \Bigl(
-\frac{1}{\euv\eir} +\frac{1}{\eir} \ln \frac{-\delta_2}{\mu} -\frac{1}{\euv}
\ln \frac{\mu}{n\cdot p}
 -\frac{1}{2}\ln^2 \frac{\mu^2}{-\delta_2 n\cdot p} -\frac{\pi^2}{12} 
\Bigr) \nonumber \\
&&+\frac{1}{(1-z)_+} \Bigl( \frac{1}{\euv} -\ln \frac{-\delta_2}{\mu}
+\ln \frac{\mu}{n\cdot p} \Bigr) -\Bigl(\frac{\ln (1-z)}{(1-z)} \Bigr)_+\Bigr], 
\nonumber \\
M_c^{(0)} &=&0,
\end{eqnarray}
where $M_c^{(0)}$ here is again subleading and is set to zero. 
The final result for the final-state jet function with the zero-bin subtraction
is written as
\begin{eqnarray}
\tilde{M}_a &=& M_a -M_a^{(0)}   \\
&=& \frac{\alpha_s C_F}{2\pi} \delta (1-z) \Bigl( \frac{1}{\euv} -\frac{1}{\eir} \Bigr)
\Bigl(\frac{1}{\euv} +\ln \frac{\mu}{n\cdot p} +1\Bigr), \nonumber \\
\tilde{M}_b &=& M_b -M_b^{(0)} \nonumber \\
&=& \frac{\alpha_s C_F}{2\pi} \Bigl\{ \delta (1-z) \Bigl[\frac{1}{\eir} \Bigl(
\frac{1}{\euv} +\ln \frac{\mu}{n\cdot p} +1\Bigr)  
+\frac{1}{\euv} \ln \frac{\mu}{n\cdot p}
+2\ln \frac{\mu}{n\cdot p} \nonumber \\
&&  +\frac{1}{2} \ln^2 \frac{\mu^2}{(n\cdot p)^2}
+2-\frac{\pi^2}{4} \Bigr]
-\frac{1}{(1-z)_+} \Bigl(\frac{1}{\euv} +1+2\ln \frac{\mu}{n\cdot p} \Bigr)
+\Bigl(\frac{\ln (1-z)}{(1-z)} \Bigr)_+ \Bigr\}. \nonumber 
\end{eqnarray}
Therefore the final-state jet function in DIS at one loop is written as
\begin{eqnarray} \label{jj}
J^{(1)} (z) &=& 2   (\tilde{M}_a +\tilde{M}_b) +M_c -\frac{\alpha_s C_F}{4\pi}
\Bigl(\frac{1}{\euv}-\frac{1}{\eir}\Bigr) \nonumber \\
&=&\frac{\alpha_s C_F}{2\pi} \Bigl\{ \delta (1-z) \Bigl[ \frac{2}{\euv^2} +
\frac{1}{\euv} \Bigl( \frac{3}{2} +2 \ln \frac{\mu^2}{Q^2}\Bigr) +\frac{3}{2}
\ln \frac{\mu^2}{Q^2} +\ln^2 \frac{\mu^2}{Q^2} +\frac{7}{2} -\frac{\pi^2}{2}
\Bigr] \nonumber \\
&&-\frac{1}{(1-z)_+} \Bigl( \frac{2}{\euv} +2 \ln \frac{\mu^2}{Q^2}  +\frac{3}{2}
\Bigr) +2 \Bigl(\frac{\ln (1-z)}{(1-z)} \Bigr)_+ \Bigr\},  
\end{eqnarray} 
where the last term in the first line comes from the virtual correction with 
the self-energy for the fermion field. In Eq.~(\ref{jj}), $n\cdot p$ is replaced by
$Q$, appropriate in DIS. This result is the same as the one in Ref.~\cite{Idilbi:2007ff}.
It is confirmed that the final-state jet function has only UV divergences after the 
zero-bin subtraction.

\section{The soft kernels $W$  and the renormalization 
group equation\label{facone}}
Combining all the one-loop results, the factorization formulae in Eqs.~(\ref{scetfac})
and (\ref{kernelw}) can be explicitly presented. The kernel $W$  to one loop is
obtained by plugging all the one-loop results in  Eqs.~(\ref{defwdy}), (\ref{defwdis}) and
(\ref{defwhig}). They are  
given as
\begin{eqnarray} \label{wone}
W_{\mathrm{DY}}  (z,\mu) &=& 
\delta (1-z) \Bigl[ 1+\frac{\alpha_s C_F}{\pi}
\Bigl(\frac{1}{\euv^2} +\frac{2}{\euv} \ln \frac{\mu}{Q} +2
\ln^2 \frac{\mu}{Q} -\frac{\pi^2}{4}\Bigr) \Bigr] \nonumber \\
&&+ \frac{\alpha_s C_F}{\pi} \Bigl[ \frac{1}{(1-z)_+} \Bigl( -\frac{2}{\euv}
-4 \ln \frac{\mu}{Q}\Bigr) +4\Bigl( \frac{\ln (1-z)}{1-z}\Bigr)_+ \Bigr],
\nonumber \\
W_{\mathrm{Higgs}}  (z,\mu) &=& 
\delta (1-z) \Bigl[ 1+\frac{\alpha_s C_A}{\pi}
\Bigl(\frac{1}{\euv^2} +\frac{2}{\euv} \ln \frac{\mu}{Q} +2
\ln^2 \frac{\mu}{Q} -\frac{\pi^2}{4}\Bigr) \nonumber \\
&&+ \frac{\alpha_s C_A}{\pi} \Bigl[ \frac{1}{(1-z)_+} \Bigl( -\frac{2}{\euv}
-4 \ln \frac{\mu}{Q}\Bigr) +4\Bigl( \frac{\ln (1-z)}{1-z}\Bigr)_+ \Bigr],
\nonumber \\
W_{\mathrm{DIS}}  (z,\mu)&=& \delta (1-z). 
\end{eqnarray}
As claimed, the soft kernels are free of IR divergences. These kernels can be computed 
systematically using perturbation theory, and the anomalous dimensions can be obtained
from Eq.~(\ref{wone}) to describe the scaling behavior.

Remarkably the kernel for DIS becomes $\delta (1-z)$ to one loop 
since $S^{(1)}_{\mathrm{DIS}} + K_{qq}^{(1)}=0$. Therefore
the inclusive DIS cross section near the endpoint consists of the hard part, the 
final-state jet function and the PDF $\phi_{q/N}$ at the scale $\mu <E$. 
It means that the soft contribution
and the initial-state jet function cancel. Therefore
the PDF satisfies the ordinary evolution equation, as in full QCD. If we do not combine
the soft function and the initial-state jet function as was done in traditional factorization
approach, the soft function and the 
collinear part  $f_{q/N}=K_{qq}\otimes \phi_{q/N}$ include IR and mixed divergences.
Then neither the soft function nor the collinear distribution function is physical, and it is 
meaningless to consider the evolution of the collinear distribution function. The solution
to this problem is our approach. That is, we combine the soft function and the initial-state
jet function to make an IR-finite kernel, which has no radiative corrections for DIS, 
and the remaining collinear part is the PDF $\phi_{q/N}$ gives the same result as in 
full QCD, given by Eq.~(\ref{phione}).  The difficulty in treating the evolution of the
collinear part without the reorganization has been discussed in Refs.~\cite{Chay:2005rz,Idilbi:2007ff,Fleming:2012kb}.

The fact that $W_{\mathrm{DIS}}=\delta (1-z)$ is  true to all orders in $\alpha_s$, 
and it can be shown by a simple argument. 
The soft zero-bin contribution from $f_{q/N}$
is obtained by integrating out the momenta of order $Q \lambda$. 
This amounts to attaching a soft gluon to a collinear fermion, and making a loop
with the soft gluon. This is exactly the procedure to obtain the eikonal
form of the soft Wilson line and to calculate its loop
correction. That is, the soft zero-bin contribution is the same as the soft function 
in DIS. 
On the other hand, the usoft zero-bin contributions vanish in $\phi_{q/N}$
to all orders in $\alpha_s$. 
It is explicitly verified here at one loop, but if we look at Eqs.~(\ref{softdy}) and 
(\ref{softdis}), the derivative term in the delta function is much smaller than $(1-z)$,
hence can be neglected. Then the delta function can be pulled out, and the remaining
usoft Wilson lines cancel. The usoft function becomes $\delta (1-z)$ 
to all orders in $\alpha_s$.  Therefore  in the absence of the usoft contributions, the 
initial-state jet function, which is the negative value of the 
soft zero-bin contributions, always cancels the soft function to all orders in $\alpha_s$
in DIS. 

In contrast, it is different in DY processes since the soft function involves
an interaction between $n$ and $\overline{n}$-collinear fermions, while the collinear 
part which interacts only within each collinear sector
does not produce the same soft interaction in the zero-bin limit.  
Note, however,  that $W_{\mathrm{DY}}$ is still IR finite though the 
soft function involves the interaction between different collinear parts,
and the initial-state jet function includes the interaction only in each collinear sector. 
The disparity between the soft function and the initial-state
jet function becomes acute in multijet processes, and it will be interesting to see if
the corresponding kernel will still remain IR finite in a more general case with multijets.

The factorized forms of the structure functions involving the kernels $W$ have
 been already shown in Eqs.~(\ref{fdyour}) and (\ref{f1our}). Each factorized 
function has a nontrivial UV behavior, however when we combine all together, 
the structure functions and the scattering cross section should have no scale dependence. 
\begin{equation} 
\label{sfs} 
\mu \frac{d}{d\mu} F_{DY} (\tau) = 0,~~~\mu \frac{d}{d\mu} F_{1} (x) = 0,
 ~~~  \mu \frac{d}{d\mu} \sigma_{\mathrm{Higgs}} =0.
\end{equation} 
Since all the elements in the factorization theorem are computed to one loop, 
the evolution of each quantity can be derived to next-to-leading logarithm 
accuracy.

The Wilson coefficients $C_{\mathrm{DIS}}(Q^2,\mu)$ and $C_H (Q,\mu)$ are  given by
\begin{eqnarray}
C_{\mathrm{DIS}} (Q^2,\mu) 
&=&1+\frac{\alpha_s C_F}{4\pi}\Bigl (-\ln^2  \frac{\mu^2}{Q^2} -3
\ln \frac{\mu^2}{Q^2} -8+\frac{\pi^2}{6}\Bigr), \nonumber \\
C_H (Q,\mu) &=&1+\frac{\alpha_s C_A}{4\pi} \Bigl(-\ln^2 \frac{\mu^2}{Q^2} 
+\frac{7}{6} \pi^2-2i\pi \ln \frac{\mu^2}{Q^2}\Bigr).
\end{eqnarray}
Using the relation $C_{\mathrm{DY}} (Q^2,\mu) =C_{\mathrm{DIS}}
(-Q^2,\mu)$, and from Eq.~(\ref{ch}), the hard coefficients to  one loop are given by
\begin{eqnarray} \label{hard}
H_{\mathrm{DIS}} (Q, \mu)&=& 1+\frac{\alpha_s C_F}{2\pi}
\Bigl (-\ln^2  \frac{\mu^2}{Q^2} -3 \ln \frac{\mu^2}{Q^2} -8
+\frac{\pi^2}{6}\Bigr), \nonumber \\
H_{\mathrm{DY}} (Q, \mu)&=& 1+\frac{\alpha_s C_F}{2\pi}
\Bigl (-\ln^2  \frac{\mu^2}{Q^2} -3 \ln \frac{\mu^2}{Q^2} -8
+\frac{7\pi^2}{6}\Bigr), \nonumber \\
H_{\mathrm{Higgs}} (Q,\mu) &=& 1+ \frac{\alpha_s C_A}{2\pi} 
\Bigl(-\ln^2 \frac{\mu^2}{Q^2} +\frac{7}{6}\pi^2\Bigr).
\end{eqnarray}

From Eq.~(\ref{hard}), the anomalous dimensions of the hard functions  in
DY, DIS and the Higgs production processes at one loop are given by 
\begin{eqnarray}\label{gh}
\gamma_{H_{\mathrm{DY}}}  (\mu) &=& 
\mu \frac{d}{d\mu} H_{\mathrm{DY}} = - \frac{\alpha_s C_F}{\pi} 
\Bigl(2 \ln \frac{\mu^2}{Q^2} + 3 \Bigr) =\gamma_{H_{DIS}}, \nonumber \\ 
\gamma_{H_{\mathrm{Higgs}}} (\mu) &=& 
\mu \frac{d}{d\mu} H_{\mathrm{Higgs}} = - \frac{2\alpha_s C_A}{\pi} 
\ln \frac{\mu^2}{Q^2}.  
\end{eqnarray} 
In the Higgs production, there is also $|C_t (m_t,\mu)|^2$ in $\sigma_0$, which 
is proportional to $\alpha_s^2$. To one loop, the anomalous dimension for
 $|C_t (m_t,\mu)|^2$ is given by
\begin{equation}
\gamma_{C_t} = \frac{d\ln |C_t|^2}{d\ln \mu} = -\frac{\alpha_s C_A}{\pi} 
\Bigl(\frac{11}{3} -\frac{2n_f}{3N_c}\Bigr), 
\end{equation}
which comes from the running of $\alpha_s$ at one loop
\begin{equation}
\mu\frac{\partial \alpha_s}{\partial \mu} =-\frac{\alpha_s^2}{2\pi} \beta_0 
=-\frac{\alpha_s^2}{2\pi}C_A \Bigl(\frac{11}{3} -\frac{2n_f}{3N_c}\Bigr).
\end{equation}

For the soft kernel $W_{\mathrm{DY}}$, $W_{\mathrm{Higgs}}$,   
and the PDF $\phi_{q/N}$, $\phi_{g/N}$,   they satisfy 
the renormalization group equation
\begin{eqnarray}
\mu \frac{d}{d\mu} W (x,\mu) &=&\int_x^1 \frac{dz}{z}\gamma_W (z,\mu) 
W\Bigl(\frac{x}{z},\mu\Bigr), \nonumber \\
\mu \frac{d}{d\mu} \phi_{q/N} (x,\mu) &=&\int_x^1 \frac{dz}{z} \gamma_q
  (z,\mu)  \phi_{q/N}\Bigl(\frac{x}{z},\mu\Bigr), \nonumber \\
\mu \frac{d}{d\mu} \phi_{g/N} (x,\mu) &=&\int_x^1 \frac{dz}{z} \gamma_g
  (z,\mu)  \phi_{g/N}\Bigl(\frac{x}{z},\mu\Bigr).
\end{eqnarray}
The anomalous dimensions $\gamma_W (z,\mu)$ and $\gamma_{q,g}
  (z,\mu) $ are schematically given as
\begin{eqnarray}
\mu \frac{d}{d\mu} Z (x,\mu) =- \int_x^1 \frac{dz}{z} 
\gamma (z,\mu) Z \Bigl(\frac{x}{z},\mu\Bigr),
\end{eqnarray}
where $Z$ is the counterterm and $\gamma$ is the anomalous dimension 
for $W$ or $\phi_{q/N}$, $\phi_{g/N}$. As discussed, 
$W_{\mathrm{DIS}}$ does not evolve.
 
From Eq.~(\ref{wone}), the anomalous dimension of the kernel, 
$W_{\mathrm{DY}}$, is given to one loop by 
\begin{equation} 
\label{gwdy} 
\gamma_{W_{\mathrm{DY}}}(x,\mu) = \frac{\alpha_s C_F}{\pi} 
\Bigl[2 \ln\frac{\mu^2}{Q^2} 
\delta(1-x) - \frac{4}{(1-x)_+} \Bigr],
\end{equation} 
and the anomalous dimension $\gamma_{W_{\mathrm{Higgs}}}$ for 
$W_{\mathrm{Higgs}}$ is obtained from $\gamma_W$
by replacing $C_F$ by $C_A$. From Eq.~(\ref{phione}), the anomalous dimensions for 
the quark PDF $\phi_{q/N}$ and the gluon PDF are given by 
\begin{eqnarray}  \label{gphi} 
\gamma_q (x,\mu) &=& \frac{\alpha_s C_F}{\pi} \Bigl[\frac{3}{2} 
\delta(1-x) + \frac{1+x^2}{(1-x)_+} \Bigr] = \frac{\alpha_s}{\pi} 
P_{qq} (x) = \frac{\alpha_s}{\pi} P_{\bar{q}\bar{q}} (x),  \nonumber \\
\gamma_g (x,\mu) &=& \frac{\alpha_s}{\pi} 2N_c 
\Bigl[ \Bigl(\frac{11}{12} -\frac{n_f}{6N_c}\Bigr) \delta (1-x) + x(1-x) +\frac{x}{(1-x)_+}\Bigr]
=\frac{\alpha_s}{\pi} P_{gg} (x), 
\end{eqnarray} 
where $P_{qq}$, $P_{\bar{q}\bar{q}}$ and $P_{gg}$ are the splitting kernels appearing
in the DGLAP evolution equations for the PDF. Here we express the anomalous dimension
before taking the limit $x\rightarrow 1$ to compare with the result in full QCD, but it is understood
that the limit $x\rightarrow 1$ should be taken near the endpoint.
The anomalous dimension for the final-state jet function is computed as 
\begin{equation} 
\label{gjdis} 
\gamma_{J}(x,\mu) = \frac{\alpha_s C_F}{\pi} \Bigl[
\Bigl( 2 \ln\frac{\mu^2}{Q^2}+\frac{3}{2}\Bigr) \delta(1-x) - 
\frac{2}{(1-x)_+} \Bigr]+ \mathcal{O} (\alpha_s^2).   
\end{equation} 

For each process, the sums of the anomalous dimensions near the endpoint $x\rightarrow 1$ 
are given as
\begin{eqnarray}
\gamma_{H_{\mathrm{DIS}}} \delta (1-x)+\gamma_q (x) + \gamma_J (x) &=& 0, 
\nonumber  \\
\gamma_{H_{\mathrm{DY}}} \delta (1-x) + \gamma_{W_{\mathrm{DY}}}(x) +2\gamma_q (x)
&=&0 ,\nonumber \\ 
(\gamma_{C_t} +\gamma_{H_{\mathrm{Higgs}}})\delta (1-x) 
 +\gamma_{W_{\mathrm{Higgs}}} (x) +2\gamma_g (x) &=&0.
\end{eqnarray}
Combining all these anomalous dimensions, we can see explicitly that 
Eq.~(\ref{sfs}) holds true to one loop near the endpoint. 

\section{Conclusion\label{conc}} 
 The traditional factorization theorems have been successful in the sense that the effects of
strong interactions at various stages have been satisfactorily separated to express
scattering cross sections as convolutions of the high-energy part, the collinear and
the soft parts. But there has remained a problem since the divergence structure is
so intricate that the collinear and the soft parts still contain UV, IR, and mixed 
divergences. Now we have better understanding of the origins of the divergences, and
this problem is taken care of by separating the soft modes consistently in the collinear part.
As a result, the factorized parts no longer involve problematic IR or mixed divergences.
With our new factorization scheme, factorization theorems have gained stronger
grounds for the theoretical description of high-energy scattering.
 
 Our factorization formula starts with a physical idea that the soft and collinear modes
should be separated at higher loops, and employs the zero-bin subtraction to realize
this idea.  Our new factorization theorem
emphasizes consistent separation of each mode at higher loops. 
In loop calculations, the soft 
contribution always encroaches on the collinear sector since the collinear loop 
momentum covers the soft region. To ensure the separation of the collinear and soft
parts, the soft contribution in the collinear sector should be consistently removed from
the collinear sector. Otherwise the overlap is bound to occur in every collinear loop
calculation. 

 The consistent treatment of the zero-bin subtraction results in the appropriate
divergence behavior. Without including the zero-bin contributions from the collinear
part, the soft contribution contains not only the IR divergence but also the mixing
of the IR and UV divergences.  The mixed divergence is especially troublesome and 
it should be absent for the soft function to have physical meaning. The inclusion 
of the initial-state jet functions $K_{qq}$ or $K_{gg}$ removes this mixed divergence. In addition,
it also changes the IR divergence to the UV divergences, and the kernels $W$
are physically meaningful. If we naively put $\epsilon =\euv =\eir$ and identify 
the poles in $\epsilon$ as the UV poles, the soft function $S_{\mathrm{DY}}^{(1)}$ in 
Eq.~(\ref{sdyone}) is identical to the kernel $W_{\mathrm{DY}}$ in Eq.~(\ref{wone}). 
It would be a good mnemonic to identify the UV divergence, but physically
it does not make sense. Since we know the origin of the divergences, the UV and IR
divergences can be systematically identified, and a physical quantity should be free of
IR divergence.

Let us finally summarize the recipe for our factorization procedure. First, we
write down the scattering cross section. Second, as in the  traditional approach, 
it can be factorized into the hard, collinear and soft parts. In full QCD, they can be
scattering amplitudes. In SCET, the hard part is obtained from the Wilson coefficient, 
the collinear and the soft parts are defined as the matrix elements of the relevant operators. 
Third, radiative corrections are computed. If there is an intermediate scale, 
the collinear distribution function with the soft zero-bin subtraction above
the scale  can be related to the PDF with the usoft zero-bin subtraction below the scale.
The relation is expressed in terms of the initial-state jet function.
Finally, we combine the soft function and the initial-state jet function to yield the soft 
kernel, which is IR finite.

Our factorization scheme gives a consistent field theoretic treatment of the UV
and IR divergences. This scheme has also been successfully applied to DY processes
with small transverse momentum \cite{Chay:2012mh}. In this case the size 
of the transverse momentum offers the intermediate scale which distinguishes 
\one\ and \two.  The initial-state jet function takes the form
\begin{equation}
f_{q/N} (x,\mathbf{k}_{\perp},\mu) =  \int_x^1 \frac{dz}{z} K_{qq}^T 
(z,\mathbf{k}_{\perp},\mu) \phi_{q/N} \Bigl(\frac{x}{z},\mu\Bigr),
\end{equation}
where $f_{q/N} (x,\mathbf{k}_{\perp},\mu) $ is the 
transverse-momentum-dependent collinear distribution function, and 
$K_{qq}^T (\mathbf{k}_{\perp},\mu)$ is the initial-state 
transverse-momentum-dependent jet function. Combining the initial-state jet function
with the transverse-momentum-dependent soft function also yields an IR-finite soft 
kernel. Therefore our factorization formalism can be applied to various high-energy
processes. It remains to be seen whether this can be a general formalism for 
factorization proof. A research in this direction is in progress.

\appendix

\section{Soft functions with  offshellness}
In the appendices, the soft functions and the collinear distribution function are computed
in the regularization scheme with the offshellness of the external particles for the 
IR divergence. By computing the soft kernel and the PDF in this scheme, we also show the
cancellation of the IR and mixed divergences.

For DIS, the soft function is defined as
\begin{equation}
S_{\mathrm{DIS}} (z) = \frac{1}{N_c} \langle 0| \mathrm{tr} \Bigl[ Y_n^{\dagger}
\tilde{Y}_{\bar{n}} \delta \Bigl(1-z +\frac{\overline{n}\cdot \mathcal{R}}{Q} \Bigr)
\tilde{Y}_{\bar{n}}^{\dagger} Y_n \Bigr] |0\rangle.
\end{equation}
But here instead of Eq.~(\ref{softwilson}) for the soft Wilson lines, the offshellness of the 
collinear particle from which the soft Wilson lines stem is inserted, and the soft Wilson lines
are modified as 
\begin{eqnarray} \label{offy}
Y_n &=& \sum_{\mathrm{perm}} \exp \Bigl[\frac{1}{n\cdot \mathcal{R} -\Delta_1 +i0}
(-g n\cdot A_s)\Bigr], \nonumber \\
  Y_n^{\dagger} &=& \sum_{\mathrm{perm}} \exp 
\Bigl[-g n\cdot A_s \frac{1}{n\cdot \mathcal{R}^{\dagger} +\Delta _1 -i0}\Bigr], \nonumber \\
\tilde{Y}_{\bar{n}} &=& \sum_{\mathrm{perm}} 
\exp \Bigl[\frac{1}{\overline{n} \cdot \mathcal{R}
+\Delta_2 -i0} (-g \overline{n}\cdot A_s )\Bigr], \nonumber \\
\tilde{Y}_{\bar{n}}^{\dagger}  &=& \sum_{\mathrm{perm}} \exp \Bigl[ -g \overline{n}\cdot A_s
\frac{1}{\overline{n}\cdot \mathcal{R}^{\dagger}-\Delta_2 +i0} \Bigr].
\end{eqnarray}
The offshellness is given by $\Delta_1 =-p_1^2/\overline{n}\cdot p_1$, and $\Delta_2 =
-p_2^2/n\cdot p_2$ where $p_1$ ($p_2$) is the $n$ ($\overline{n}$) collinear momentum
of the collinear particles with the corresponding soft Wilson lines to be attached.
Note that the insertion of $\Delta_1$ and $\Delta_2$ looks similar to the rapidity
regulator, but it is the regulator for the IR divergence. Though 
similar in form, their sources are distinct. If we put the offshellness explicitly, 
$\Delta_i$ and $\delta_i$ take 
different forms. In the soft Wilson line, $\Delta_i$ can be obtained from the offshellness
of a single collinear particle where the soft gluons are attached. On the other hand, the
rapidity regulator $\delta_i$ is obtained by the emission of the $n$-collinear gluons from all the
collinear or heavy particles in other directions.  Therefore $\delta_i$ have complicated
dependence on the offshellness of the other particles. Only in the back-to-back
current, there exists a simple relation $\delta_1 = \Delta_2$, $\delta_2=\Delta_1$.  

In obtaining the dependence on the offshellness, we consider collinear particles or 
antiparticles from $-\infty$, or to $\infty$, as considered in Ref.~\cite{Chay:2004zn},
and assign nonzero offshellness to the collinear particles.  
Here also arises the problem of gauge invariance. But it suffices to say that this is only an 
intermediate step to regulate IR divergences with the offshellness 
since the dependence of the offshellness is cancelled in the final results.

Here we employ the dimensional regularization for the UV divergence, and the IR divergence
appears as logarithms of $\Delta_1$ or $\Delta_2$. 
The virtual correction in Fig.~\ref{softdisf} (a), and the real gluon emission in 
Fig.~\ref{softdisf} (b) are given as
\begin{eqnarray}
M_{s,\mathrm{DIS}}^a 
&=& -2ig^2 C_F \delta (1-z) \dimf \int \dfl \frac{1}{l^2(n\cdot l -\Delta_1) 
(\overline{n}\cdot l -\Delta_2 )} \nonumber \\
&=&-\frac{\alpha_s C_F}{2\pi} \delta (1-z) \Bigl( \frac{1}{\epsilon^2} 
+\frac{1}{\epsilon}\ln \frac{\mu^2}{\Delta_1 \Delta_2} +\frac{1}{2}
\ln^2 \frac{\mu^2}{\Delta_1 \Delta_2}+\frac{\pi^2}{4}\Bigr), \nonumber \\
M_{s,\mathrm{DIS}}^b &=& 4\pi g^2 C_F \dimf \int \dfl \frac{1}{(n\cdot l +\Delta_1)
(\overline{n}\cdot l+\Delta_2)} \delta (l^2)  \delta \Bigl( 1-z-\frac{\overline{n}\cdot l}{Q}
\Bigr) \nonumber \\
&=&\frac{\alpha_s C_F}{2\pi} \Bigl[ \delta (1-z) \Bigl( -\frac{1}{\epsilon} \ln 
\frac{\Delta_2}{Q} -\ln \frac{\Delta_2}{Q}\ln \frac{\mu^2}{-p_1^2} +\frac{1}{2}
\ln^2 \frac{\Delta_2}{Q} +\frac{\pi^2}{6} \Bigr) \nonumber \\
&&+\frac{1}{(1-z)_+} \Bigl( \frac{1}{\epsilon} +\ln \frac{\mu^2}{-p_1^2} \Bigr) 
-\lone \Bigr].
\end{eqnarray}
In calculating these matrix elements, the following plus distribution functions are
used.
\begin{eqnarray}
\frac{1}{1-z+\delta} &=& -\delta (1-z) \ln \delta +\frac{1}{(1-z)_+}, \nonumber \\
\frac{\ln (1-z)}{1-z+\delta}&=& \delta (1-z) \mathrm{Li}_2 \Bigl(-\frac{1}{\delta} \Bigr)
+\Bigl(\frac{\ln (1-z)}{1-z}\Bigr)_+, 
\end{eqnarray}
where $\mathrm{Li}_2 (x)$ is the dilogarithmic function. 
The one-loop result for the soft function in DIS is given by
\begin{eqnarray} \label{msdis}
S_{\mathrm{DIS}}^{(1)} (z) &=& 2 \mathrm{Re}\,
(M_{s,\mathrm{DIS}}^a  +M_{s,\mathrm{DIS}}^b) 
\nonumber \\
&=& \frac{\alpha_s C_F}{\pi} \Bigl[\delta (1-z) \Bigl( -\frac{1}{\epsilon^2} -\frac{1}{\epsilon}
\ln \frac{\mu^2}{-p_1^2} -\frac{1}{2} \ln^2 \frac{\mu^2}{-p_1^2}-\frac{\pi^2}{12}
\Bigr) \nonumber \\
&&+\frac{1}{(1-z)_+} \Bigl(\frac{1}{\epsilon} +\ln \frac{\mu^2}{-p_1^2} \Bigr)  -\lone\Bigr].
\end{eqnarray}
Note that the soft function does not depend on $\Delta_2$, as it should be. 
In DIS, the final state is described by the final-state jet function which depends on $\Delta_2$,
but $\Delta_2$ is not the IR cutoff. Instead it is related to the invariant jet mass of the
final states. Therefore $\Delta_2$ does not represent the IR divergence, and the soft function
is independent of $\Delta_2$.

For DY process, the soft function is defined as
\begin{equation}
S_{\mathrm{DY}} (z) = \frac{1}{N_c} \langle 0| \mathrm{tr} \Bigl[Y_n^{\dagger}
Y_{\bar{n}} \delta \Bigl( 1-z +\frac{2v\cdot \mathcal{R}}{Q}\Bigr) Y_{\bar{n}}^{\dagger}
Y_n\Bigr]|0\rangle.
\end{equation}
The soft Wilson lines $Y_{\bar{n}}$ and $Y_{\bar{n}}^{\dagger}$ are defined in the 
same way as $Y_n$ and $Y_n^{\dagger}$ except that $n$ is replaced by $\overline{n}$,
and $\Delta_1$ by $\Delta_2$.

The virtual correction and the real gluon emission are given as
\begin{eqnarray} \label{msdy}
M_{s,\mathrm{DY}}^a &=&  -2ig^2 C_F \delta (1-z) 
\dimf \int \dfl \frac{1}{l^2 (n\cdot l -\Delta_1)
(\overline{n}\cdot l +\Delta_2)}   \\
&=&-\frac{\alpha_s C_F}{2\pi} \delta (1-z) \Bigl( \frac{1}{\epsilon^2}
+\frac{1}{\epsilon} \ln \frac{\mu^2}{\Delta_1 (-\Delta_2)} +\frac{1}{2}
\ln^2 \frac{\mu^2}{\Delta_1 (-\Delta_2)} +\frac{\pi^2}{4}\Bigr),\nonumber \\
M_{s,\mathrm{DY}}^b &=& 4\pi g^2 C_F \dimf \int \dfl
\frac{1}{(n\cdot l +\Delta_1)(\overline{n}\cdot l-\Delta_2)} \delta (l^2) \delta \Bigl(
1-z-\frac{2v\cdot l}{Q}\Bigr) \nonumber \\
&=& \frac{\alpha_s C_F}{2\pi} \Bigl[ \delta (1-z) \Bigl(\ln \frac{\Delta_1}{Q} 
\ln \frac{-\Delta_2}{Q} -\frac{\pi^2}{6}\Bigr) -\frac{1}{(1-z)_+} 
\ln \frac{\Delta_1(- \Delta_2)}{Q^2}
+2\lone\Bigr]. \nonumber
\end{eqnarray}
In the final stage of computing $M_{s,\mathrm{DY}}^b$, there are two possibilities in taking
the limit $\Delta_1, \Delta_2 \rightarrow 0$.  That is, the limit 
$\Delta_1 \rightarrow 0$ can be approached first with $\Delta_2$ fixed, and then the limit
$\Delta_2 \rightarrow 0$ is taken. The limiting procedure can be reversed, however, the 
result is the same irrespective of the order of taking limits.

The soft function in DY process obtained by adding the hermitian conjugate of 
Eq.~(\ref{msdy}), and is given by
\begin{eqnarray} \label{sdyoff}
S_{\mathrm{DY}}^{(1)} (z) &=& 2 \mathrm{Re} \, (M_{s,\mathrm{DY}}^a +M_{s,\mathrm{DY}}^b)
\\
&=& \frac{\alpha_s C_F}{\pi} \Bigl\{ \delta (1-z) \Bigl[ -\frac{1}{\epsilon^2}
-\frac{1}{\epsilon} \Bigl( \ln \frac{\mu^2}{-p_1^2} + \ln \frac{\mu^2}{-p_2^2}
-\ln \frac{\mu^2}{Q^2}\Bigr)   -\frac{5\pi^2}{12} 
\nonumber \\
&&-\frac{1}{2} \ln^2 \frac{\mu^2}{-p_1^2}
-\frac{1}{2} \ln^2 \frac{\mu^2}{p_2^2} +\frac{1}{2} \ln^2  \frac{\mu^2}{Q^2}\Bigr]
\nonumber \\
&&+\frac{1}{(1-z)_+} \Bigl( \ln \frac{\mu^2}{-p_1^2} + \ln \frac{\mu^2}{-p_2^2}
-2\ln \frac{\mu^2}{Q^2}\Bigr) +2\lone\Bigr\}. \nonumber
\end{eqnarray}

As can be seen again in Eqs.~(\ref{msdis}) and (\ref{msdy}), the soft functions contain
IR divergences as well as mixed divergences. Therefore the soft functions themselves 
are not physical.

\section{Collinear distribution functions with  offshellness}
The collinear distribution functions can also be evaluated with the offshellness 
for the IR regulator. The poles in $\epsilon$ are UV divergences.
The naive collinear matrix elements are given as
\begin{eqnarray}
M_a &=& \frac{\alpha_s C_F}{2\pi} \delta (1-z) \Bigl[ \frac{1}{\epsilon} 
\Bigl( 1+\ln \frac{\delta}{Q}\Bigr) + \ln \frac{\mu^2}{-p^2} -\frac{1}{2}
\ln^2  \frac{\delta}{Q} + \ln \frac{\mu^2}{-p^2} \ln \frac{\delta}{Q} 
+2-\frac{\pi^2}{3}\Bigr],  \nonumber  \\
M_b &=& \frac{\alpha_s C_F}{2\pi}  \Bigl[ \delta (1-z) \Bigl( -\frac{1}{\epsilon} 
\ln \frac{\delta}{Q} -\ln \frac{\delta}{Q} \ln \frac{\mu^2}{-p^2} 
+\frac{1}{2} \ln^2 \frac{\delta}{Q} +\frac{\pi^2}{6}\Bigr) \nonumber \\
&&+\frac{z}{(1-z)_+} \Bigl(\frac{1}{\epsilon} -\ln z + \ln \frac{\mu^2}{-p^2} 
\Bigr) -z \Bigl(\frac{\ln (1-z)}{(1-z)} \Bigr)_+\Bigr], \nonumber \\
M_c &=&\frac{\alpha_s C_F}{2\pi} (1-z) \Bigl( \frac{1}{\epsilon} +
\ln \frac{\mu^2}{-p^2} -2 -\ln z(1-z)\Bigr).  
\end{eqnarray}
Here we put  $\overline{n}\cdot p =Q$, and $\delta$ is the rapidity regulator in the 
collinear Wilson line. If $\delta$ is not used, IR poles in $\eir$ appear instead, and they
also cancel.

The soft zero-bin contributions are given as
\begin{eqnarray}
M_a^{(0)} &=& -\frac{\alpha_s C_F}{2\pi} \delta (1-z) \Bigl[  \frac{1}{\epsilon^2}
+\frac{1}{\epsilon} \ln \frac{\mu^2 Q}{-p^2\delta}
+\frac{1}{2} \ln^2 \frac{\mu^2 Q}{-p^2\delta} +\frac{\pi^2}{4}\Bigr], \nonumber \\
M_{b,\mathrm{s}}^{(0)} &=& \frac{\alpha_s C_F}{2\pi}  \Bigl[\delta (1-z) \Bigl(
-\frac{1}{\epsilon} \ln \frac{\delta}{Q} -\ln \frac{\delta}{Q} \ln \frac{\mu^2}{-p^2}
+\frac{1}{2} \ln^2 \frac{\delta}{Q} +\frac{\pi^2}{6} \Bigr)   \nonumber \\
&&+\frac{1}{(1-z)_+} \Bigl(\frac{1}{\epsilon} +\ln \frac{\mu^2}{-p^2} \Bigr) -\lone
\Bigr], \nonumber \\
M_c^{(0)} &=&0.
\end{eqnarray}

Therefore the collinear contributions with the soft zero-bin subtractions are given as
\begin{eqnarray}
\tilde{M}_a &=& M_a -M_a^{(0)} \nonumber \\
&=& \frac{\alpha_s C_F}{2\pi} \delta (1-z) \Bigl[ \frac{1}{\epsilon^2} 
+\frac{1}{\epsilon}\Bigl( 1+ \ln \frac{\mu^2}{-p^2}  \Bigr) + \ln \frac{\mu^2}{-p^2} 
+\frac{1}{2} \ln^2 \frac{\mu^2}{-p^2} +2-\frac{\pi^2}{12}\Bigr], \nonumber \\
\tilde{M}_{b,\mathrm{s}} &=& M_b -M_{b,\mathrm{s}}^{(0)} \nonumber \\
&=&\frac{\alpha_s C_F}{2\pi}  \Bigl[    - \Bigl(\frac{1}{\epsilon}
+\ln\frac{\mu^2}{-p^2}\Bigr) 
-\frac{z \ln z}{(1-z)_+}
+\ln (1-z) \Bigr], \nonumber \\
\tilde{M}_c &=& M_c = \frac{\alpha_s C_F}{2\pi} (1-z) \Bigl( \frac{1}{\epsilon} +
\ln \frac{\mu^2}{-p^2} -2 -\ln z(1-z)\Bigr).  
\end{eqnarray}

The collinear quark distribution function at one loop is given as
\begin{eqnarray}
f_{q/N}^{(1)} (z,\mu)&=& 2 \mathrm{Re} \,(\tilde{M}_a +\tilde{M}_b) +M_c + 
\delta (1-z) (Z_{\xi}^{(1)} + R_{\xi}^{(1)})  \\
&=& \frac{\alpha_s C_F}{2\pi}  \Bigl[ \delta (1-z)\Bigl( \frac{2}{\epsilon^2} 
+\frac{3}{2\epsilon}+ \frac{2}{\epsilon} \ln \frac{\mu^2}{-p^2}   +\frac{3}{2}
\ln \frac{\mu^2}{-p^2} 
+  \ln^2 \frac{\mu^2}{-p^2} +\frac{7}{2}-\frac{\pi^2}{6} \Bigr)\nonumber \\
&& -\Bigl( \frac{1}{\epsilon} +\ln \frac{\mu^2}{-p^2}\Bigr) (1+z) -\frac{1+z^2}{(1-z)_+} \ln z
-2 (1-z) +(1+z) \ln (1-z) \Bigr], \nonumber
\end{eqnarray}
where $Z_{\xi}^{(1)}$ is the counterterm and $R_{\xi}^{(1)}$ is the residue in
the self-energy of the fermion $\xi_n$ at one loop and they are given as
\begin{equation}
Z_{\xi}^{(1)} = -\frac{\alpha_s C_F}{4\pi} \frac{1}{\epsilon}, \ R_{\xi}^{(1)} 
= -\frac{\alpha_s C_F}{4\pi} \Bigl( 1+\ln \frac{\mu^2}{-p^2}\Bigr).
\end{equation}

The usoft zero-bin contribution differs only in the real gluon emission, which is given as
\begin{eqnarray}
M_{b,\mathrm{us}}^{(0)} 
&=& \frac{\alpha_s C_F}{2\pi} \delta (1-z) \Bigl[ \frac{1}{\epsilon^2}
+\frac{1}{\epsilon} \Bigl( \ln \frac{\mu^2}{-p^2} -\ln \frac{\delta}{Q}\Bigr) +\frac{\pi^2}{4}
  \\
&&+ \frac{1}{2} \ln ^2 \frac{\mu^2}{-p^2} -\ln \frac{\mu^2}{-p^2}\ln \frac{\delta}{Q}
-\frac{1}{2} \ln^2 \frac{\delta}{Q}  \Bigr]. \nonumber
\end{eqnarray}
The corresponding collinear part with the usoft zero-bin subtraction is given as
\begin{eqnarray}
\tilde{M}_{b,\mathrm{us}} &=& M_b -M_{b,\mathrm{us}}^{(0)} \\
&=& \frac{\alpha_s C_F}{2\pi} \Bigl[ \delta (1-z) \Bigl( -\frac{1}{\epsilon^2}
-\frac{1}{\epsilon} \ln \frac{\mu^2}{-p^2} -\frac{1}{2} \ln^2 \frac{\mu^2}{-p^2} 
-\frac{\pi^2}{12}\Bigr) \nonumber \\
&&+\frac{z}{(1-z)_+} \Bigl( \frac{1}{\epsilon} +\ln \frac{\mu^2}{-p^2} -\ln z\Bigr)
-z \lone \Bigr]. \nonumber 
\end{eqnarray}
And the one-loop correction to the PDF is given as
\begin{eqnarray}
\phi_{q/N}^{(1)}  (z,\mu) &=& \frac{\alpha_s C_F}{2\pi} \Bigl[ \delta (1-z) \Bigl(
\frac{3}{2\epsilon} +\frac{3}{2}\ln \frac{\mu^2}{-p^2} +\frac{7}{2}-\frac{\pi^2}{3} \Bigr) 
+\Bigl( \frac{1}{\epsilon} +\ln \frac{\mu^2}{-p^2} \Bigr) \frac{1+z^2}{(1-z)_+} 
\nonumber \\
&&-2 (1-z) - (1+z^2) \frac{\ln z}{(1-z)_+} -(1+z^2) \lone\Bigr].
\end{eqnarray}

The initial-state jet function can be written as
\begin{eqnarray}
K_{qq} (z,\mu) &=& 2 \mathrm{Re}\,(-M_{b,\mathrm{s}}^{(0)} +M_{b,\mathrm{us}}^{(0)}) \\
&=& \frac{\alpha_s C_F}{\pi} \Bigl[ \delta (1-z)\Bigl( \frac{1}{\epsilon^2} +\frac{1}{\epsilon}
\ln \frac{\mu^2}{-p^2} +\frac{1}{2} \ln^2 \frac{\mu^2}{-p^2} +\frac{\pi^2}{12} \Bigr)
\nonumber \\
&&-\Bigl(\frac{1}{\epsilon} +\ln \frac{\mu^2}{-p^2} \Bigr) \frac{1}{(1-z)_+} +\lone\Bigr].
\nonumber
\end{eqnarray}
To one loop, the kernel $W$ are given as
\begin{eqnarray} \label{woff}
W_{\mathrm{DIS}}^{(1)} &=& 0,  \nonumber \\
W_{\mathrm{DY}}^{(1)} &=& \frac{\alpha_s C_F}{\pi} \Bigl[\delta (1-z) \Bigl(
\frac{1}{\epsilon^2} +\frac{1}{\epsilon} \ln \frac{\mu^2}{Q^2} +\frac{1}{2} \ln^2
\frac{\mu^2}{Q^2} +\frac{\pi^2}{4}  \Bigr)\nonumber \\
&&-\Bigl(\frac{2}{\epsilon}  
+2\ln \frac{\mu^2}{Q^2} \Bigr) \frac{1}{(1-z)_+} +4 \lone\Bigr].
\end{eqnarray}
In this calculational scheme, the kernels $W$ are also IR finite.
As in the case with the dimensional regularization, $W_{\mathrm{DIS}}^{(1)} =0$. However,  
$W_{\mathrm{DY}}^{(1)}$ is the same as the result with the dimensional regularization
except the $\pi^2$ term. The  source of the disparity can be seen from Eq.~(\ref{sdyoff}). 
There is a term $\ln^2 (\mu^2 /p_2^2)$ in $S_{\mathrm{DY}}$, while there is
$\ln^2 (\mu^2 /-p_2^2)$ in $K_{qq}$. Therefore the IR divergence cancels as expected, but
there is a remnant of $\pi^2$. It results in the difference of the term with $\pi^2$.  
If we take the limit $p_1^2, p_2^2 \rightarrow 0$, it corresponds to the dimension regularization
limit for the IR divergence. Then the logarithms turn into poles in $\eir$, and there is no 
additional $\pi^2$ term involved. However, if we strictly keep the signs of the offshellness, 
this additional factor of $\pi^2$
appears. If there were only single logarithms, this ambiguity does not occur.
We have not been able to confirm whether the difference is due to the scheme
dependence and further consideration is needed.

\acknowledgments
J.~Chay  and C.~Kim were supported by Basic Science Research Program 
through the National Research Foundation of Korea (NRF)
funded by the Ministry of Education, Science and Technology 
(No. 2012R1A1A2008983 and No. 2012R1A1A1003015),  respectively.

\end{document}